\documentclass[screen, acmsmall]{acmart}
\usepackage[utf8]{inputenc}
\usepackage[T1]{fontenc}
\usepackage{blindtext, graphicx}
\usepackage[english]{babel}
\usepackage{algorithm,algpseudocode}
\usepackage{amsmath}
 
\usepackage{amsfonts}
\usepackage{amssymb}
\usepackage{wrapfig}
\usepackage{lipsum}
\usepackage{caption}
\usepackage{subcaption}
\usepackage{pgffor}
\usepackage{enumerate}
\usepackage{enumitem} 
\usepackage{babel}
\usepackage{balance}
\usepackage{chemgreek,textgreek}
\usepackage{pifont}
\usepackage{enumitem}
\usepackage{graphicx,wrapfig,lipsum}
\usepackage{hyperref}
\usepackage{multirow}
\usepackage{multicol}
\usepackage{todonotes}
\usepackage{color}
\usepackage{xcolor}
\usepackage{colortbl}
\usepackage{booktabs}
\usepackage{flushend}

\def\BibTeX{{\rm B\kern-.05em{\sc i\kern-.025em b}\kern-.08emT\kern-.1667em\lower.7ex\hbox{E}\kern-.125emX}}

\begin{document}

\acmJournal{PACMHCI}
\acmYear{2019} \acmVolume{3} \acmNumber{CSCW}
\acmArticle{151} \acmMonth{11} \acmPrice{15.00}
\acmDOI{10.1145/3359253}

\received{April 2019}
\received[revised]{June 2019}
\received[accepted]{August 2019} 
%
\title[Modeling Islamist Extremist Communications on Social Media]{Modeling Islamist Extremist Communications on Social Media using Contextual Dimensions: Religion, Ideology, and Hate}

%

\author[Kursuncu]{Ugur Kursuncu}
\affiliation{%
  \institution{Kno.e.sis Center, Wright State University}
  \city{Dayton}
  \state{OH}
  \country{USA}}
\email{ugur@knoesis.org}

\author[Gaur]{Manas Gaur}
\affiliation{%
  \institution{Kno.e.sis Center, Wright State University}
  \city{Dayton}
  \state{OH}
  \country{USA}}
\email{manas@knoesis.org}

\author[Castillo]{Carlos Castillo}
\affiliation{%
  \institution{Universitat Pompeu Fabra}
  \city{Barcelona}
  \country{Spain}}
\email{chato@acm.org}

\author[Alambo]{Amanuel Alambo}
\affiliation{%
  \institution{Kno.e.sis Center, Wright State University}
  \city{Dayton}
  \state{OH}
  \country{USA}}
\email{amanuel@knoesis.org}

\author[Thirunarayan]{Krishnaprasad Thirunarayan}
\affiliation{%
  \institution{Kno.e.sis Center, Wright State University}
  \city{Dayton}
  \state{OH}
  \country{USA}}
\email{t.k.prasad@wright.edu}

\author[Shalin]{Valerie Shalin}
\affiliation{%
  \institution{Department of Psychology, Wright State University}
  \city{Dayton}
  \state{OH}
  \country{USA}}
\email{valerie.shalin@wright.edu}

\author[Achilov]{Dilshod Achilov}
\affiliation{%
 \institution{Department of Political Science, University of Massachusetts, Dartmouth}
 \state{MA}
 \country{USA}}
\email{dachilov@umassd.edu}

\author[Arpinar]{I. Budak Arpinar}
\affiliation{%
  \institution{Department of Computer Science, The University of Georgia}
  \city{Athens}
  \state{GA}
  \country{USA}}
\email{budak@uga.edu}

\author[Sheth]{Amit Sheth}
\affiliation{%
  \institution{AI Institute, The University of South Carolina}
  \city{Columbia}
  \state{SC}
  \country{USA}}
\email{amit@sc.edu}

%
\renewcommand{\shortauthors}{Ugur Kursuncu et al.}


%
\begin{abstract}
Terror attacks have been linked in part to online extremist content. Online conversations are cloaked in religious ambiguity, with deceptive intentions, often twisted from mainstream meaning to serve a malevolent ideology. Although tens of thousands of Islamist extremism supporters consume such content, they are a small fraction relative to peaceful Muslims. The efforts to contain the ever-evolving extremism on social media platforms have remained inadequate and mostly ineffective. Divergent extremist and mainstream contexts challenge machine interpretation, with a particular threat to the precision of classification algorithms. Radicalization is a subtle \emph{long-running persuasive} process that occurs over time. 
Our context-aware computational approach to the analysis of extremist content on Twitter breaks down this persuasion process into building blocks that acknowledge inherent \emph{ambiguity} and \emph{sparsity} that likely challenge both manual and automated classification. Based on prior empirical and qualitative research in social sciences, particularly political science, we model this process using a combination of three \emph{contextual dimensions} -- religion, ideology, and hate -- each elucidating a degree of radicalization and highlighting independent features to render them computationally accessible. We utilize domain-specific knowledge resources for each of these contextual dimensions such as Qur'an for \emph{religion}, the books of extremist ideologues and preachers for political \emph{ideology} and a social media hate speech corpus for \emph{hate}. The significant sensitivity of the Islamist extremist ideology and its local and global security implications require reliable algorithms for modelling such communications on Twitter. Our study makes three contributions to reliable analysis: (i) Development of a computational approach rooted in the contextual dimensions of  religion, ideology, and hate, which reflects strategies employed by online Islamist extremist groups, (ii) An in-depth analysis of relevant tweet datasets with respect to these dimensions to exclude likely mislabeled users, and (iii) A framework for understanding online radicalization as a process to assist counter-programming. Given the potentially significant social impact, we evaluate the performance of our algorithms to minimize mislabeling, where our context-aware approach outperforms a competitive baseline by 10.2\% in precision, thereby enhancing the potential of such tools for use in human review. 
\end{abstract}


%


%

%

\maketitle

\section{Introduction}
\label{sec:intro}
In December, 2018 the United Nations Counter-Terrorism Implementation Task Force (CTITF)\footnote{\url{https://www.un.org/counterterrorism/ctitf/en/about-task-force }} met\footnote{\url{https://www.un.org/sg/en/content/sg/speeches/2018-12-06/un-global-counter-terrorism-compact-coordination-committee-remarks}} in New York focusing on key action items for a global effort to understand the distortions of the narratives used by terrorists on online platforms. Their subsequent report emphasized that ``terrorist organizations like Da'esh (ISIS) and Al Qaida continue to twist religion to serve their ends. The  threat posed by returning and relocating fighters, as well as from individuals inspired by them, remains high and has a global reach''. Approximately one thousand Americans between 1980 and 2011 and more than five thousand individuals from Europe through 2015, \emph{traveled} to join extremist groups abroad \cite{boutin2016foreign}. Since 2011, the Federal Bureau of Investigation (FBI) reported that 300 Americans attempted or traveled to Syria and Iraq to join extremist groups \cite{meleagrou2018travelers}. Since March, 2014 at least 182 individuals have been charged in the US for ISIS-related offenses\footnote{GW Extemism Tracker: ISIS in America (May 2019) \url{https://extremism.gwu.edu/isis-america}}. Recent reports \cite{frampton2017new} suggest that the terror attacks were linked  online extremist content, consumed by supporters as the newly radicalized recruits living in the West are active users of Twitter (e.g., ISIS supporters had higher activity than 67\% of all Twitter users) \cite{berger2015isis}. Recently, a 24 year old college student from Alabama became radicalized on Twitter before moving to Syria to join ISIS\footnote{\url{https://www.nytimes.com/2019/02/22/podcasts/the-daily/isis-american-women.html }}\footnote{\url{https://www.nytimes.com/2019/02/19/us/islamic-state-american-women.html}}. Her radicalization began when she was 20 through meeting other Muslim community members on Twitter, which  she refers to as the \emph{Muslim twittersphere}. Self-taught, she read verses from the Qur'an but  interpreted them with others in the twittersphere, persuaded that when \emph{the true Islamic State} is declared, it is obligatory to do \emph{hijrah}, which they see as the pilgrimage to \emph{'the Islamic State'}. The lack of adequate knowledge about the religion combined with the  extremist ideology on Twitter led her to regrettable decisions and actions. 

Much has been written on how effectively violent terrorist networks, most notoriously ISIS, have utilized social media to recruit new members \cite{vidino2015isis}. Nevertheless, efforts to capture systematically the ever-evolving dynamics of extremism on social media platforms have remained inadequate: limited in scope, opaque in approach, and mostly ineffective in practice\footnote{\url{https://www.lawfareblog.com/marginalizing-violent-extremism-online }}\footnote{\url{https://www.theguardian.com/world/2017/sep/19/britain-has-large-audience-for-online-jihadist-propaganda-report-says }}\footnote{\url{https://policyexchange.org.uk/wp-content/uploads/2017/09/The-New-Netwar-2.pdf}}\cite{alava2017youth, de2017radicalisation, hussain2014jihad}. Further, as Islamist extremism is a subjective concept with serious repercussions for individuals, the analysis of this topic imposes significant social responsibility in designing reliable algorithms, to avoid discriminatory or biased classification. Merely knowing that someone is a Muslim should not label him or her as a religious  extremist. Therefore, domain expertise and responsible use of knowledge provides decisive context.

In this study, we model such religious  extremist communications on Twitter based on highly persuasive content, incorporating domain-specific knowledge with three distinct  contextual dimensions: religion (\textbf{R}), extremist Islamist ideology (\textbf{I}) and hate (\textbf{H}), originating from domain expert's analysis of the data (see Section \ref{sec:eda}) as well as social science literature \cite{van2003measurement, loza2007psychology, schafer2002spinning} in consultation with our domain expert co-author. The \textbf{religion} dimension refers to Muslim attitudes that range from ``mainstream'' through more ``extreme'' interpretations of Islamic scriptures. Attitudes toward political \textbf{extremist ideology} (i.e., Islamism) are another prevalent dimension of extremism. The conceptualization and measurement of variations in political and ideological attitudes toward Islamism are drawn from \cite{achilov2017got}'s (concept building) study of Political Islamism. Finally, \textbf{hate speech} or attitudinal support for violence is the third critical dimension that provides a benchmark for Islamist extremism with the potential for violent terrorist acts \cite{helfstein2012edges,hafez2015radicalization}. Our approach will further enable an analysis of the radicalization process individually and collectively in a fine-granular manner, to provide a computational foundation for any form of human intervention. Our hypothesis is that the combination of these three contextual dimensions will create  more coherent and distinctive representation of extremist communications, improving the performance of classification. Accordingly, we address the following research questions: \textbf{RQ1:} Does incorporation of these contextual dimensions into representation of social media communications improve extremist content classification performance? \textbf{RQ2:} Which combination of the dimensions are more effective? \textbf{RQ3:} How much does each of these dimensions contribute to the classifier performance?

To achieve these goals, we perform an in-depth analysis of datasets that contain \emph{verified} Islamist extremist accounts (see Section \ref{sec:D} for details). We operationalize abstract models of behavior exhibited by an extremist individual under the influence of religion, extremist ideology, and hate. We generate representations for different contextual dimensions using word embeddings drawn from domain-specific resources, to render these dimensions computationally accessible. We then address the challenging problems of inherent sparsity and ambiguity in relations that are implicit in this data to obtain reliable results. Further, language and topical analyses characterize the similarity between users that we can scrutinize, and hierarchical clustering identifies outlier individuals that otherwise would mislead the analysis. Finally, we model Islamist extremist communications utilizing supervised classification algorithms operating over domain-specific representations of extremist and non-extremist users generated by incorporating the three contextual dimensions mentioned above. 

In this pursuit, our study makes the following four specific contributions: (i) Development of a computational approach rooted in the dimensions of religion, extremist ideology, and hate that are employed by online Islamist extremist groups to influence, (ii) an in-depth data analysis of relevant datasets with respect to these dimensions, demonstrating improvement in classification with ideological and hate contents, (iii) a framework for understanding online radicalization that serves as a basis for counter programming, (iv) potentially significant reduction in discriminatory bias of mislabeling mainstream (non-extremist) Muslim accounts as extremist\footnote{Mainstream adherents to Islam number $\sim$1.8 billion, while only  a small fraction of them adopt extremist views. \\ \url{https://www.pewresearch.org/fact-tank/2017/08/09/muslims-and-islam-key-findings-in-the-u-s-and-around-the-world/}}. As precision in classifying the communications on Islamist extremism takes precedence over recall to interpret a response properly, we evaluate precision to emphasize minimizing mislabeled non-extremist users. Our approach outperforms a competitive baseline by 10.2\% in precision and 8.5\% in recall leading to an improvement of 10.7\% in F1-score. The combination of all three contextual dimensions in the representation of an account outperforms other alternatives.

In Section \ref{sec:relatedwork}, we provide details on existing research related to Islamist extremism on social media. In Section \ref{sec:preliminaries}, we describe the preliminary concepts used in this study. Section \ref{sec:eda} characterizes the dataset through language, topical, statistical and similarity analyses, which inform our subsequent modeling approach. We discuss the detailed modeling of Islamist extremism in Section \ref{sec:method}, and the evaluation of results and its implications in Section \ref{sec:results}. Finally, we present our conclusions with future directions in Section \ref{sec:conclusion}.

\section{Related Work}
\label{sec:relatedwork}
The state of the art approaches to detecting and analyzing Islamist extremist communications on social media are limited in their selection of features due to the sparsity and ambiguity inherent in social media data. We need novel approaches to learn coherent representations of content by making use of contextually relevant domain knowledge in a principled manner.
 
Previous research related to Islamist extremism on social media has focused on four problems: (i) detection of extremist content \cite{saif2017semantic, kaati2015detecting, arpinar2016social}, (ii) prediction of extremist users \cite{fernandez2018understanding, fernandez2018contextual, ferrara2016predicting, rowe2016mining, wadhwa2013tracking, anwar2015ranking}, (iii) detection of communities for extremist users \cite{ashcroft2015detecting, scanlon2014automatic, scanlon2015forecasting, agarwal2015open} and (iv) identification of hate promoting extremism \cite{cano2013weakly, agarwal2014focused, agarwal2016spider, agarwal2015using, sureka2014learning}. We categorize this study as detection of extremist content and prediction of extremist users.

Ferrara et al. \cite{ferrara2016predicting} proposed a framework to predict extremist users, their adoption of extremist content, and interaction reciprocity between extremists and regular users. They built predictive models for a binary classification of extremist users using  Random Forest (RF) and Logistic Regression (LR) algorithms. To predict extremist users, they employed 52 features that include user and tweet metadata as well as information related to user network and  temporal evolution of content. They performed prediction of adoption of extremist viewpoints (as a result of being influenced) based on the behavior of regular users retweeting the content from extremist users. The prediction of interactions with extremists (indicating more active involvement) was based on reply tweets. For all three prediction tasks, RF outperformed LR, with AUCs (Area Under the ROC Curve) of 0.87, 0.77 and 0.69 for the prediction of extremist users, adoption and interactions, respectively.

Rowe et al. \cite{rowe2016mining} performed an analysis of 154K Twitter users in order to extract cues related to radicalization from their content, based on whether the users favor pro vs anti-extremist stances. They found that 727 of these users displayed a pro-ISIS stance, particularly when an event related to ISIS unfolded. In another study, a graph-based semantic approach for detection of radicalization in the Twitter content was proposed by \cite{saif2017semantic}. They utilized knowledge graphs (e.g., DBpedia) to provide semantic relationships between the extracted entities in the content, improving robustness over prior approaches that involved lexical, sentiment, topic and network features. They applied their approach to 1132 (566 pro / 566 anti-ISIS) users, with 1.9M (0.6M pro / 1.3M anti-ISIS) tweets, achieving an F1-score of 0.92. 

Fernandez et al. \cite{fernandez2018understanding} developed an approach for detection and prediction of the influence a user is exposed to, by combining social and computational models of radicalization. They compared the radicalization level of 112 pro-ISIS v/s 112 ``general'' Twitter users with respect to the roots of radicalization at the individual (micro), community (meso) and global (macro) levels. Their approach achieved up to a 0.90 F1-score for detection and between 0.70 and 0.80 precision for prediction, utilizing vector representations of users designed based on their three level approach. In a follow-up study, Fernandez et al. \cite{fernandez2018contextual} utilized contextual semantic features of radical content on social media employing ontologies and knowledge bases (DBpedia and Wikidata) to capture categories, topics, entities and entity types. They tested the effectiveness of extracting semantic context from radical conversations on classification of pro-ISIS and non pro-ISIS accounts. They achieved an improvement in precision, recall and F1-score of 0.04, 0.04 and 0.03, respectively.

In contrast to this literature, our work is grounded in the social science literature  and incorporates domain-specific resources \cite{sheth2017knowledge, gaur2018let, gaur2019knowledge} in the model to better understand and detect extremist content using linguistic approaches. As Islamist extremism is a complex issue that involves different contexts, traditional approaches do not adequately capture important nuances in the language related to the multiple contextual dimensions of the problem. Our approach uncovers these nuances by decomposing social media posts along the three contextual dimensions (see Section \ref{sec:CDM}), and provides a fine-grained basis for understanding an individual's progression towards radicalization. Moreover, this understanding will improve interpretability and serve as a crucial basis for designing and building counter extremism narratives for possible de-radicalization efforts.


\section{Preliminaries}
\label{sec:preliminaries}
\subsection{Background: Islamist Extremism On Social Media}
Extremist actors involved in the dissemination of persuasive content frequently disguise themselves as legitimate  representatives of a religion, doctrine or ideology (e.g., extremists posing as true (mainstream) believers in Islam). From the perspective of the persuader, persuasive (propagandist) messages should resemble messages produced by common agents, but be able to perpetuate their hidden agenda by deception and foster misinformation by distorting concepts and relations. This challenges the reliable detection of radicalization content. Such persuasive content involves unconstrained doctrinal concepts and relationships inspired by religion, history and politics.

For example, the concept ``jihad'' commonly appears in mainstream Islamic as well as extremist communications, albeit with different context-dependent interpretations (see Table \ref{tab:examples}). The concept of ``jihad'' can mean (i) self-spiritual struggle, (ii) defensive war to protect lives and property from aggression, or (iii) acts of provoked or unprovoked violence, depending on its context of use \cite{cook2015understanding}. Classification of the first and the second interpretation of ``jihad'' as extreme would cause the computational model to be gravely incorrect. Further, the degree and progression on a radicalization scale are reflected in the content. For example, users who are recruiters will have a tendency to disseminate information to influence/impress their followers, and initially utilize religious references in their narratives. As they move further in their persuasive radicalization process, they use extremist ideology propaganda by referring to resources of their ideologues. The process culminates with inciting violence by utilizing hate speech and encouraging the followers to act and commit violence. Thus, to glean reliable and comprehensive insights and to assess its intensity, it is critical to use the three contextual dimensions of Religion(R), Ideology(I) and Hate(H) to analyze all  communication.

\begin{table}[t]
\footnotesize
\begin{center}
\begin{tabular}{p{0.5cm}|p{11cm}p{0.25cm}p{0.25cm}p{0.25cm}}
    \toprule[2.5pt]
     \textbf{No.} & \textbf{Extremist Content Examples} & \textbf{R} & \textbf{I} & \textbf{H} \\ \midrule[1pt]
     1. & ``Here is the fragrance of \textbf{Paradise}, Here is the field of \textit{Jihad}. Here is the \textit{land of \#Islam}, Here is the \textit{land of the Caliphate}'' & & \checkmark & \\ 
     2. & ``Reportedly, a number of \textit{apostates} were \underline{killed} in the process. Just because they like it I guess. \textit{\#SpringJihad} \underline{\#CountrysideCleanup}'' & & & \checkmark \\
     3. & ``and \textit{Jihad} means to \textit{sacrifice} YOURSELF in \underline{war} to save your \underline{country} (or \textbf{religion})'' & & \checkmark & \checkmark \\
     4. & ``I asked about the paths to \textbf{Paradise} It was said that there is no path shorter than \textit{Jihad}'' & \checkmark & \checkmark & \\
     5. & ``\textbf{God} honored us w/ \textit{Jihad Khilafah} in this era of \textit{Fitnah}'' & & \checkmark & \\
     6. & ``By the \textbf{Lord of Muhammad} (blessings and \textbf{peace be upon him}) The \textit{nation} of \textit{Jihad} and \textit{martyrdom} can never be \textit{defeated}'' & & \checkmark & \\
     7. & ``Anyone who prefers to raise \textit{secularism} over \textbf{Islam} is a \textit{kafir}, whether he’s from Saudi, Sudan, Somalia, Mexico, Burma, Hawaii, or elsewhere.'' & & \checkmark & \\
     8. & ``\textbf{\#MyJihad} is to take care of mother, then mother, then mother, then father, then other relatives in...'' & \checkmark & & \\
     9. & ``Kindness is a language which the blind can see and the deaf can hear \textbf{\#MyJihad} be kind always'' & \checkmark & & \\
     10. & ``May \textbf{Allah} accept those who \textbf{fast} Monday's and Thursday's.'' & \checkmark & & \\
     \bottomrule[2.5pt]
\end{tabular}
\end{center}
\caption{\footnotesize Example tweets from our dataset for extremist/non-extremist social media users, annotated by our co-author domain expert for religion (R), extremist ideology (I) and hate (H) terminology. ``Jihad'' appears in multiple dimensions. Examples 8 and 9 contain the term ``Jihad'' in its mainstream meaning, whereas, in Examples 1, 2, 3, 4, 6, it refers to its meaning in the extremist context. Some of the terms are coded based on their relatedness to one of the three dimensions: Religion (Bold-faced), Ideology (Italicized) and Hate (Underlined)}
\label{tab:examples}
\vspace{-3.0em}
\end{table}

As the interpretation of an individual term depends upon its surrounding lexical context, accurate identification of the relationship between lexical features and Islamist extremism is crucial. For example (see also  Table \ref{tab:examples}), when the term ``jihad'' co-occurs with ``kill'' and ``attack'', it connotes \textit{hate and violence}. In the presence of ``Allah'' and ``Islam'', the term ``jihad'' stands for its original meaning denoting the \textit{religious} concept of self-struggle. ``Jihad'' co-occuring with ``imam\_anwar\_al\_awlaki'', who is considered \cite{bowman2012exploring} a prominent \textit{ideologue} of radical Islamist groups, connotes \textit{exhorting hate and violence}. The word ``jihad'' acquires a different meaning in each of the contexts above; therefore, its representation should be semantically different as well. For this reason, we generate representations of contents and users based on the three contextual dimensions (Religion, Extremist Ideology and Hate) that we identified for the domain of Islamist extremism. The representation of contents is created through word embedding models learned from  domain-specific resources, for each contextual dimension. We provide further details of these procedures in the subsequent subsections. 

\subsection{Dataset}
\label{sec:D}
Our ground truth dataset includes 538 extremist users and their 47,376 tweets in the positive class spanning nearly seven years between October 2010 and August 2017. We have used two datasets (i) tweets of Pro-ISIS users\footnote{\url{https://www.kaggle.com/fifthtribe/how-isis-uses-twitter}}, and (ii) tweets of users reported by the Lucky Troll Club\footnote{\url{http://archive.is/V24aS }} that have been \emph{verified and suspended by Twitter} because of ISIS-related supportive activity\footnote{\url{https://www.technologyreview.com/s/603626/data-mining-reveals-the-rise-of-isis-propaganda-on-twitter/ }}\cite{badawy2018rise}. The data and labels were manually curated by annotators who are experts in the Arabic language and verified by Twitter’s anti-abuse team. The dataset has also been used in recent studies for modeling radicalization \cite{ferrara2016predicting, fernandez2018understanding}. From this dataset, we selected only English tweets. In our positive samples, the prevalent concepts, topics and terms usually refer to key domain-specific entities such as people (e.g., ideologues, historic person), locations (region, city) and verbs (fight, kill, join). In the rest of this paper, we refer to the positive examples as ``extremist users'' and to the negative examples as ``non-extremist users''.

\paragraph{\textbf{Creation of Negative Class Samples:}} For development and testing, we use a dataset of 6040 non-extremist mainstream Muslim religious users and 7000 of their tweets created by Chen et al. \cite{chen2014us} to constitute the negative class. Note that Islamist extremist content and  mainstream/non-extremist content have overlapping vocabulary terms (e.g., jihad) though used in different senses. This word-sense disambiguation challenges the accurate detection of extremist cues in the content, reducing precision. To evaluate the effectiveness of our approach to disambiguate content, we create the negative class dataset from a dataset of Muslim religious users. We employ Hierarchical Dirichlet Processing (HDP) (non-parametric) clustering \cite{teh2005sharing} on the tweets from users to hierarchically organize users probabilistically based on their content's  topical similarity. The application of HDP over the 7K tweets resulted in 600 coherent clusters based on 20 topics with 30 sub-topics each \cite{srijith2017sub}. We treat these clusters as standing for non-extremist users, and randomly select 538 from the 600-user clusters to create our non-extremist user dataset. This approach allows us to deal with data sparsity associated with users without sacrificing their coherence from their normal usage.

\subsection{Word Embeddings}
\label{sec:WE}
Generating embeddings of content provides a numerical vector representation that captures the context of a word/phrase in a corpus. Embedding algorithms including Word2Vec \cite{mikolov2013distributed}, GLoVe \cite{pennington2014glove} and FastText \cite{athiwaratkun2018probabilistic} have proven to be effective for creating rich representations tuned to a specific domain. Word (or phrase, sentence, document) embedding models generate numerical vector representations of words (or phrases, sentences, documents) that can be used to represent the content \cite{kursuncu2019predictive}. We can use vector operations such as addition, multiplication, and concatenation to aggregate word representations into representations of phrases, sentences, or short documents such as tweets.  Domain-specific words can have frequencies, neighboring words, and usages  that differ significantly. Hence, it is important to learn embeddings based on domain-specific corpora, e.g., related to Islamist extremism. 

\subsection{Contextual Dimension Modeling}
\label{sec:CDM}

Highly persuasive religious extremist communications on Twitter frequently employ language and topical cues related to the religion of Islam (R), extremist Islamist ideology (I) and hate (H) for effectiveness (see Section \ref{sec:eda}). The distribution of prevalent terms (i.e., words, phrases, concepts) in the content of both extremist and non-extremist users reflects different contextual dimensions of the Islamist extremism problem (see Tables \ref{tab:N} and \ref{tab:T}). For example, extremist users often make references to concepts and terminologies related to extremist ideology and hate language, while they share content containing relatively fewer Islamic concepts. On the other hand, in the content of non-extremist users, they more often share content related to the religion of Islam unlike extremist users. Moreover, the ambiguity of diagnostic terms (e.g., jihad) also mandates representation of terms in different contexts.  Therefore, to better reflect these differences, we create multiple models that will represent the three contextual dimensions for a reliable analysis. We were guided by authoritative sources to ground our hypotheses and look to operationalize the approach that was not performed for social media communications by social scientists. Specifically, \cite{van2003measurement, loza2007psychology, schafer2002spinning, qin2007analyzing, awan2017cyber, bunt2003islam} show that extremism and subsequent acts of violence are linked to each of these contexts, describing that extremist groups create different interpretations of \emph{religion} that serves their political \emph{extremist ideological} interests, inciting \emph{hate} and finally leading individuals to commit acts of violence. While \cite{van2003measurement} found, with  context  theory,  significant  positive  correlations  between  cognitive  complexity and extremist  ideology, \cite{loza2007psychology} argued that extremists teach their young followers to hate "the West" forming an \emph{ideological} premise that they should follow \emph{the ‘true’ Islam} and it should be considered above everything. Further, \cite{schafer2002spinning} found that there are connections between web sites operated by \emph{extremist} and \emph{hate} organizations and select episodes of \emph{violence}. In the lights of these findings from the literature and our observations from our dataset, we identify the three contextual dimensions by carefully scrutinizing the Islamist extremist communications on Twitter as well as social science literature in consultation with our domain expert co-author. 

\begin{wrapfigure}{r}{7.5cm}
\vspace{-0.5em}
  \begin{center}
    \includegraphics[width=0.33\textwidth, 
    trim=5.0cm 2.5cm 0.5cm 0.5cm]{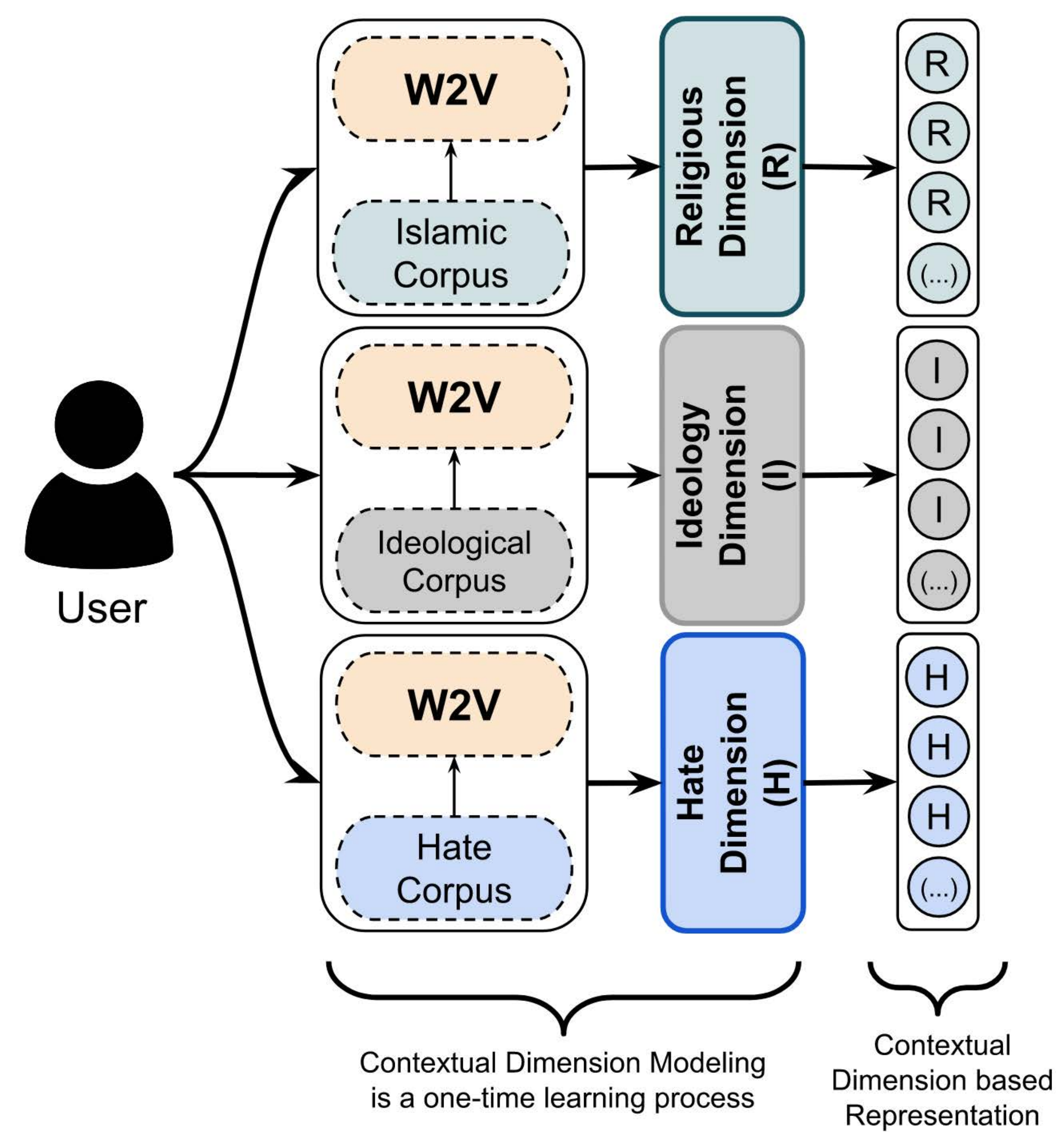}
  \end{center}
  \vspace{1em}
  \caption{\footnotesize Creation of representations for a user using contextual dimension models using Word2Vec (W2V).}
  \label{fig:DM}
\vspace{-0.5em}
\end{wrapfigure}

We create three word embedding models for the three contextual dimensions, employing \emph{domain specific resources}\footnote{These resources are available upon request to reproduce our experiments.} as follows: (i) For Religion: The Qur'an English translation\footnote{\url{https://www.noblequran.com/translation/}} by Muhammad Taqi-ud-Din Al-Hilali and Muhammad Muhsin Khan, and Hadith (collection of Prophetic Narrations) resources referenced by ISIS the most\footnote{\url{ https://www.kaggle.com/fifthtribe/isis-religious-texts }}: Sahih Al-Bukhari\footnote{\url{https://en.wikipedia.org/wiki/Sahih\_al-Bukhari }} and Sahih Muslim\footnote{\url{https://en.wikipedia.org/wiki/Sahih\_Muslim }} well-known authentic Hadith collection, (ii) For Extremist Ideology: Magazines published by ISIS (e.g., Dabiq\footnote{\url{https://en.wikipedia.org/wiki/Dabiq\_(magazine) }}, Rumiyah\footnote{\url{https://en.wikipedia.org/wiki/Rumiyah\_(magazine)}}), books and transcribed lectures of extremist ideologues identified by our domain expert co-author (e.g., Anwar Al-Awlaki, Hassan Al-Banna, Said Qutb, Yusuf al-Qaradawi, Abul A’la Maududi), and (iii) For Hate: Hate speech corpus \cite{hateoffensive}. Figure \ref{fig:DM} illustrates the overall flow of creation of contextual dimension models and representations of a user for each contextual dimension.

In this study, we use Word2Vec \cite{mikolov2013efficient} with skip-grams to generate contextual dimension models. As we aggregate the tweets for each user, we take \textit{union} of unigrams $(U)$, bigrams $(B)$, and trigrams $(T)$, from tweets of a user, and ``average'' their word embeddings to generate an embedding vector of the user, following the commonly-used method. We formally define our contextual dimension modeling procedure as follows:\\
Let us represent a set of words/phrases in tweets of a user $(u)$ as $\mathbf{G}$ which is generated through the following procedure: $(U\setminus\{B, T\}) \cup (B \setminus \{T\}) \cup T)$. Then, we generate representations of a user $(u)$ along the three contextual dimensions of Religion (R), Ideology (I), and Hate (H) as follows:
\begin{equation}
	\vec{\mathcal{V}}(u^d) = \frac{\sum_{w \in (G \cap W^d)} \vec{\mathbf{\nu}}_{w}}{|G \cap W^d|}, d \in \mathbf{\Psi}
\label{eq:1}
\end{equation}

where, $\mathbf{\Psi}$ denotes contextual dimensions, $\vec{\mathcal{V}}(u^d)$ represents the embedding vector of a user $(u)$ generated from a word embedding model for a contextual dimension \emph{d} having vocabulary $\mathbf{W^d}$, and $w$ is a word in a tweet of a user $u$ in the vocabulary $\mathbf{W^d}$. The denominator of the Equation \ref{eq:1} is the cardinality of the set of words in the intersection between tweets of a user ($\mathbf{G}$) and vocabulary of a dimension ($\mathbf{W^d}$). We generate the embedding of a user along three dimensions as follows:

\begin{equation}
    \vec{\mathcal{V}}(u) = SVD(\mathcal{V}(u^R) \oplus \mathcal{V}(u^I) \oplus \mathcal{V}(u^H))
\end{equation}
where ($\oplus$) concatenates the vector representations of a user (\emph{u}) along the three contextual dimensions: R: $\mathcal{V}(u^R)$,  I: $\mathcal{V}(u^I)$, and H: $\mathcal{V}(u^H)$. Then, we utilize singular value decomposition(SVD) to perform dimensionality reduction \cite{shin2018interpreting} which reduces the dimensions to 300, which is standard in the word embeddings literature \cite{mikolov2013efficient,pennington2014glove,bojanowski2017enriching}. 

\begin{figure}[t]
    \centering
    \includegraphics[width=0.50\textwidth, 
    trim=5.0cm 2.5cm 2.5cm 0.5cm, angle=-90]{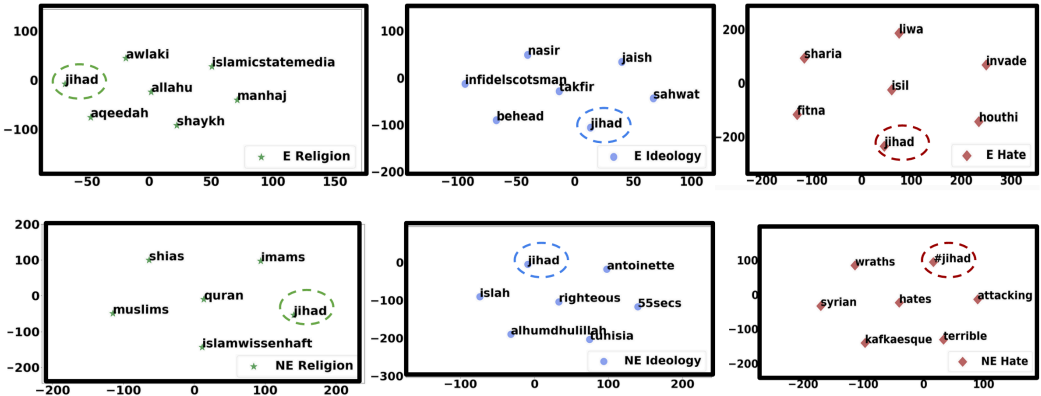}

  \vspace{-4.5em}
  \caption{\footnotesize  The representations of the word “jihad” with  different meanings based on different contextual dimensions. Upper figure: The closest terms to “jihad” in the contexts: religion, ideology and hate for \textbf{extremist} (E) users. Lower figure: The closest terms to “jihad” in the three contexts: religion, ideology and hate for \textbf{non-extremist} (NE) users.}
  \label{fig:words-2d-space}
  \vspace{-1.5em}
\end{figure}

As we argue that such representations of a user involving the three contextual dimensions will be more coherent and representative, it will disambiguate diagnostic domain-specific words/phrases providing different representations. Thus, the very same terms with different meanings will be represented differently as well. To illustrate how the meaning changes in the extremist and non-extremist content with respect to each contextual dimension, we show the representations of the term ``jihad'' and its closest terms in Figure \ref{fig:words-2d-space}, for extremist and non-extremist users. In (a), the closest terms to ``jihad'', in the content of extremist users, are related to the concepts that extremist groups use to justify their ideology. These closest terms include ``infidelscotsman'', ``behead'', ``nasir'', ``takfir'' \textit{for ideology}; ``isil'', ``fitna'', ``houthi'', ``invade'' \textit{for hate}; and ``aqeedah'', ``awlaki'', ``shaykh'', ``allahu'', \textit{for religion}. On the other hand, (b) displays  the terms closest to ``jihad'' that are mostly related to the mainstream (non-extremist) Islamic terminologies; such as ``alhamdulillah'', ``islah'', ``righteous'' \textit{for ideology}, ``quran'', ``muslims'', ``imams'' \textit{for religion}, and ``terrible'', ``attacking'', ``hates'' \textit{for hate}. As the meaning of the word ``jihad'' changes depending on its context, its numerical representation changes as well.

\section{Exploratory Data Analysis}
\label{sec:eda}
As noted, Islamist extremism on social media has security implications, and requires careful judgment and reliable labeling of content and individuals. Hence, before attempting to use our dataset for modeling Islamist extremism, we examine our dataset carefully, identifying patterns and potential anomalies, checking our assumptions, and testing our  hypothesis. We use a multi-pronged approach involving lexical, topical, statistical, and user similarity as discussed below.
\subsection{N-Gram Analysis}
\label{sec:NG}
The language characteristics of extremist and non-extremist contents differ with respect to different contextual dimensions. To determine the language characteristics, we extract n-grams (n=1 to 3) from tweets of users in extremist and non-extremist datasets using the skip n-gram model \cite{mikolov2013efficient, bouma2009normalized}. In our experiments with n-grams, we have observed that 2- and 3-grams were particularly informative because multi-word concepts and entities are prevalent in these communications. For example, ``\emph{imam anwar al awlaki}'' occurs often because he is a prominent and popular extremist ideologue. Moreover, Islamist extremist groups including ``Ahrar al-Sham'', ``Jabhat al-Nusra'', ``Islamic State'' (IS), and locations such as ``Deir ez-Zor'' once held by ISIS in the Syria and Iraq region are mentioned frequently.  

\begin{table}[t]
\footnotesize
\begin{center}
\begin{tabular}{p{2cm}p{5.5cm}p{5.5cm}}
    \toprule[2.5pt]
     \textbf{N-grams} & \textbf{Extremist Users} & \textbf{Non-Extremist Users} \\ \midrule[1pt]
     Unigrams & \textit{isis}, \textit{syria}, \underline{kill}, \textit{iraq}, \textbf{muslim, allah}, \underline{attack}, \underline{break}, \textit{aleppo, assad, islamicstate, army}, soldier, cynthiastruth, \textbf{islam}, support, \textit{mosul, libya, rebel}, \underline{destroy}, \textit{airstrike} & person, majd, opening, \textbf{belief}, follower, knowing, khazarrose, al-beltagy's, forza, rally, smyrna, togethernabilahasya, cyrus, \textbf{islam}, okuyamadigimi, dibawain, \textbf{waalaikumsalam} \\ \midrule[1pt]
     Bigrams & \textit{Caliphate news, islamic state, iraq army, soldier kill, iraqi army, syria isis, syria iraq, assad army, terror group, shia militia, isis attack, aleppo syria, martyrdom operation, ahrar sham, assad regime, follow support, lead coalition, turkey army}, isis claim, \underline{kill isis} & sleep controversial, time activist, \textbf{ali alhamduillah}, pape gratefulness, out violence, inii riots, anti-muslim fukushima's, afternoon commit, agree regimes, \#patientsafety personality, mahdi muslims, movie muslimap, ahmad worried, biblical festival, jummah soldier, mixe masjid, masmilwaukee reminder, mubarak title, imams koplok \\ \midrule[1pt]
     Trigrams & \textit{Imam anwar awlaki}, video message islamicstate, \underline{fight islamic state}, \textit{isisclaim responsibility attack}, \textit{muwahideen powerful middleeast}, isis tikrit tikritop, amaqagency islamicstate fighter, \textit{sinai explosion target, alone state fighter}, \underline{intelligence reportedly kill}, \textit{khilafahnew islamic state}, \underline{yemanqaida commander kill}, isis militant hasakah, breakingnew assad army, \underline{isis explode middle}, hater trier haleemah, trust isis tighten, \underline{qamishlus isis fighting}, \textit{defeat enemy allah}, \underline{kill terrorist baby}, \textit{ahrar sham leader} &  allah bowtie raised, holars killll studios, muhammad lingkgan fdraiser, homefeed wajib akal, israeli paid fajr, eradicating nations project, 2500 muslims homicides, suicide espinoza excess, flow producin shiekh, non-muslims defend reality, masalah taft makan, beneficial right knalan, push serious idea, jahannam philosophy prostration, brotherhood tranquility korean, \textbf{saturday defile astagfirullah}, quick taught america, \textbf{bbe quran goal}, \textbf{alhamdulillah sat week}, touching kids killed, fodation islamic state, islamicate samajhten defined \\
     \bottomrule[2.5pt]
\end{tabular}
\end{center}
\caption{\footnotesize Most prevalent unigrams, bigrams, trigrams in the content of extremist and non-extremist users. The n-grams related to the religion of Islam, extremist Islamist ideology and hate appear in bold, italics, and underlined, respectively.}
\label{tab:N}
\vspace{-3.5em}
\end{table}

Among the users in the extremist dataset, terms such as \emph{``allah'', ``fear allah'', ``jannah'', ``muslim'', ``attack'', ``kill'', ``isis'',} and \emph{``islamic state''}, are the most frequent, where the first four terms are related to the religion of Islam. The terms \emph{attack} and \emph{kill} are related to hate, and the last two terms refer to the most prominent Islamist extremist group (see Table \ref{tab:N}). In contrast, among the users in the non-extremist dataset, the most frequent n-grams are \emph{``amendment yourself'', ``\#truthmonkey'', ``booth volunteering'', ``time activist'', ``drink upon''}, which reflect their contrasting social/political communication.

\subsection{Topical Analysis}
\label{sec:TA}
\begin{wrapfigure}{r}{7cm}
  \begin{center}
    \includegraphics[width=0.32\textwidth, 
    trim=5.0cm 2.5cm 3.5cm 0.5cm]{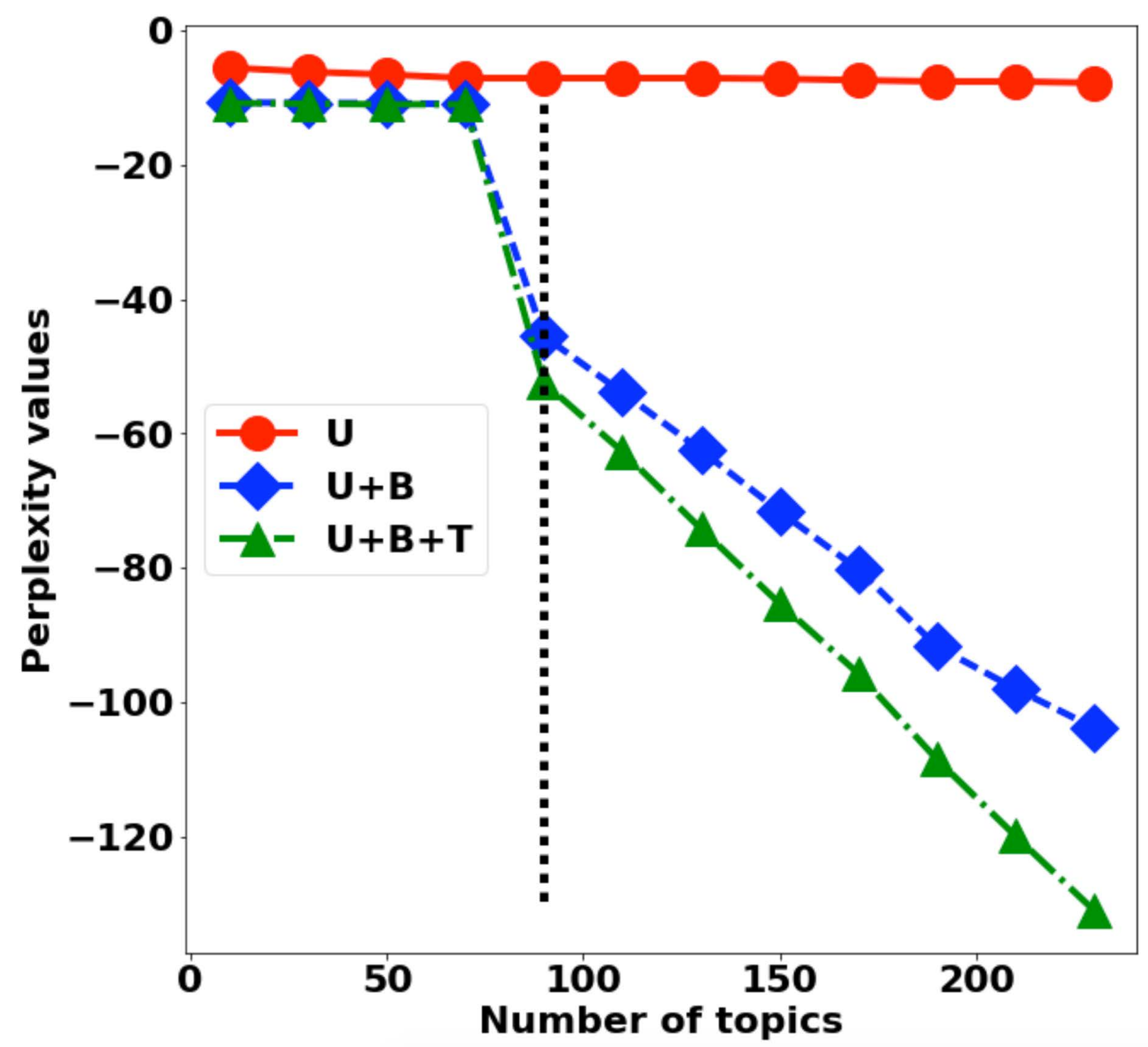}
  \end{center}
  \caption{\footnotesize Identification of optimal number of relevant topics based on perplexity scores. From these graphs, we identify 90 as an optimal number of topics that best represent the content of users.}
  \label{fig:topic-perplexity}
\end{wrapfigure}

Topics in the content of extremist and non-extremist users can play a critical role in determining intra-class similarity and inter-class differences among users. We used Latent Dirichlet Allocation (LDA) over n-grams (n=1-3) to assess topical similarity of content of \emph{extremist} users to characteristics of Islamist extremism. As LDA is a parametric probabilistic approach for retrieving topics, the number of topics should be carefully selected to capture the themes \emph{optimally}. Hence, we used \emph{perplexity measure} to obtain the optimal number of topics that best represents the content. A higher perplexity score implies higher representativeness and semantic integrity among the topics. Researchers \cite{gaur2018let} have applied LDA over various combinations of unigrams (U), bigrams (B), and trigrams (T) to obtain informative topics. We apply this procedure creating three different topic models covering the following combinations: (i) U, (ii) U+B, (iii) U+B+T (see Section \ref{sec:D}). Using the perplexity score depicted in Figure \ref{fig:topic-perplexity}, we identified 90 as the threshold for the optimal number of topics for each user for each of the four topic models. We have also used the same number (90) of topics for non-extremist users. That is, we did not compute  perplexity scores for non-extremist users separately as they were created using Hierarchical Dirichlet Process (HDP), which is a non-parametric form of LDA.

\begin{table}[!htbp]
\footnotesize
\begin{center}
\begin{tabular}{p{3cm}p{10.5cm}}
    \toprule[2.5pt]
     \textbf{Dataset (User types)} & \textbf{Prevalent Topics} \\ \midrule
     \textbf{Extremist Users} & \textit{islamic state, syria, isis}, \underline{kill}, \textbf{allah}, video, minute propaganda video scenes, \textit{jaish islam release}, \underline{restock missile}, \textit{kaffir, join isis}, aftermath, \textbf{mercy}, \textit{martyrdom operation syrian opposition, punish libya isis, syria assad}, islam sunni, swat, lose head, \textit{wilayatalfurat, somali},\underline{ child kill}, \textit{takfir, jaish fateh, baghdad, iraq}, kashmir muslim, capture, \textit{damascus}, report rebel,\textit{ british}, qala moon, \textbf{jannat}, \textit{isis capture, border cross, aleppo}, iranian soldier, tikrit tikrittop, \underline{lead shia military kill}, \textit{saleh abdeslam refuse cooperate} \\ \midrule[1pt]
     
     \textbf{Non-Extremist Users} & \underline{masjid job kill}, smelled valentines day myanmar, black pada newspaper, \textbf{quran}, tarek necessary like lost, radioactive bande khuda delivered, kaiciid united nations between sky, movement,  mustafa human reference, dislodge fatir, kids cruise islamophobia language, active people justice party, \textbf{hati tiba jihad}, abdel lawful farrakhan, adha suhaib, hiasan racist, \textbf{darinya alhamdulilah}, order u.s. iran strike, light headed narcissist stuff, truth monkey, protest jihad controversial, moon accept boycott states, arabi fornicate expiration, al-beltagy rose, \textbf{khuda jannat}, brotherhood maaf, \textbf{sunni islam, wasidiyah allahumma}, muhammad laws onward walking, desperation rather hugo, okurs show rotinhell, american smurf, abraham killed, shifters controversy military, \textbf{allah, prophet muhammad}, rest peace, iraq asks bible jerebu \\
     \bottomrule[2.5pt]
\end{tabular}
\end{center}
\caption{\footnotesize Topics extracted from the content of extremist and non-extremist users using LDA and HDP. The topics related to religion, ideology and hate are bold-faced, italicized and underlined respectively. Remaining topics did not fall under any particular dimension.}
\label{tab:T}
\vspace{-3em}
\end{table}

In the content of extremist users, topics related to the Islamist ideology are prevalent compared to topics related to hate and religion. For instance, Islamist extremist users frequently make use of ideology-related words/phrases to promote their organization (e.g., ``islamic state'', ``isis'', ``join isis''), its activities (e.g., ``martyrdom operation syrian opposition'') or attacks on non-muslim people (e.g.,  ``kaffir'') (see Table \ref{tab:T}). On the other hand, topics related to religion are prevalent in the content of non-extremist users. For instance, non-extremist users invoke religious concepts such as ``allah'', ``prophet muhammad'', ``quran'', ``alhamdulillah'' and ``jannat'', whereas there is only one reference to hate, ``masjid job kill''. We observe that the prevalence of these topics  related to different contextual dimensions in the content of extremist and non-extremist users varies.

\subsection{User Similarity}
\label{sec:US}
\begin{figure}[!htbp]
  \begin{center}
    \includegraphics[width=0.80\textwidth, 
    trim=5.0cm 2.5cm 3.5cm 0.5cm]{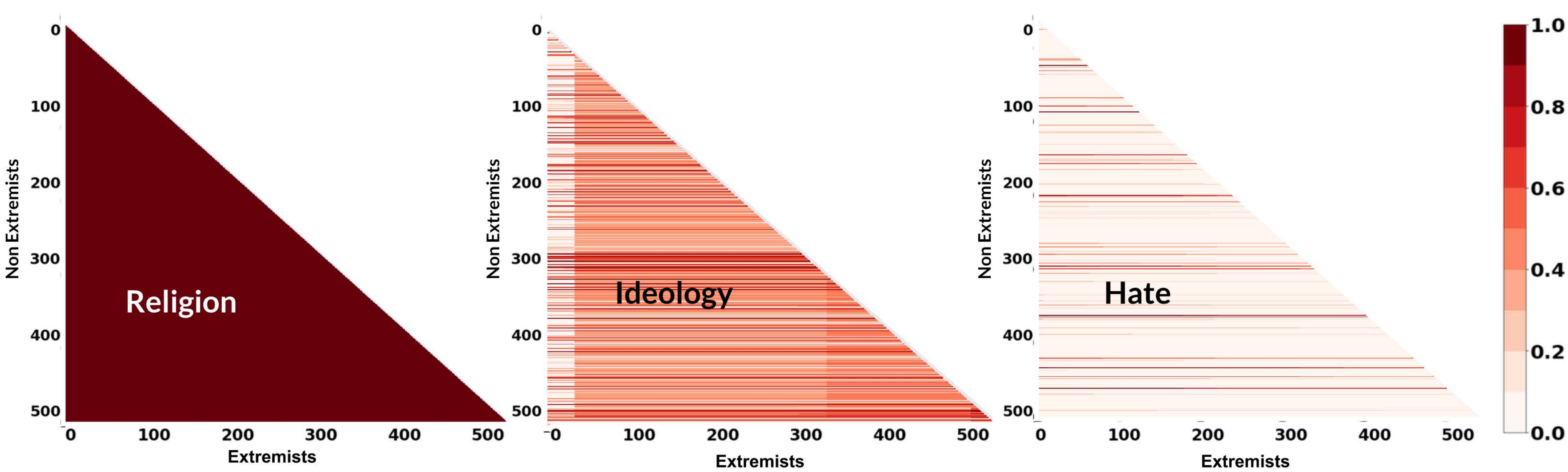}
  \end{center}
  \vspace{0.5em}
  \caption{\footnotesize Similarity between extremist (x-axis) and non-extremist users (y-axis) based on religion, ideology and hate dimensions. User id appears on the x axis for extremist and the y axis for non-extremist users. Extremist and non-extremist users show strong similarity for religion, and weak similarity for hate. They show stronger similarity based on ideology compared to hate, but weaker compared to religion.}
  \label{fig:hmap-NE}
\vspace{-0.5em}
\end{figure}

\begin{figure}[!htbp]
  \begin{center}
    \includegraphics[width=0.80\textwidth, 
    trim=5.0cm 1.5cm 3.5cm 0.5cm]{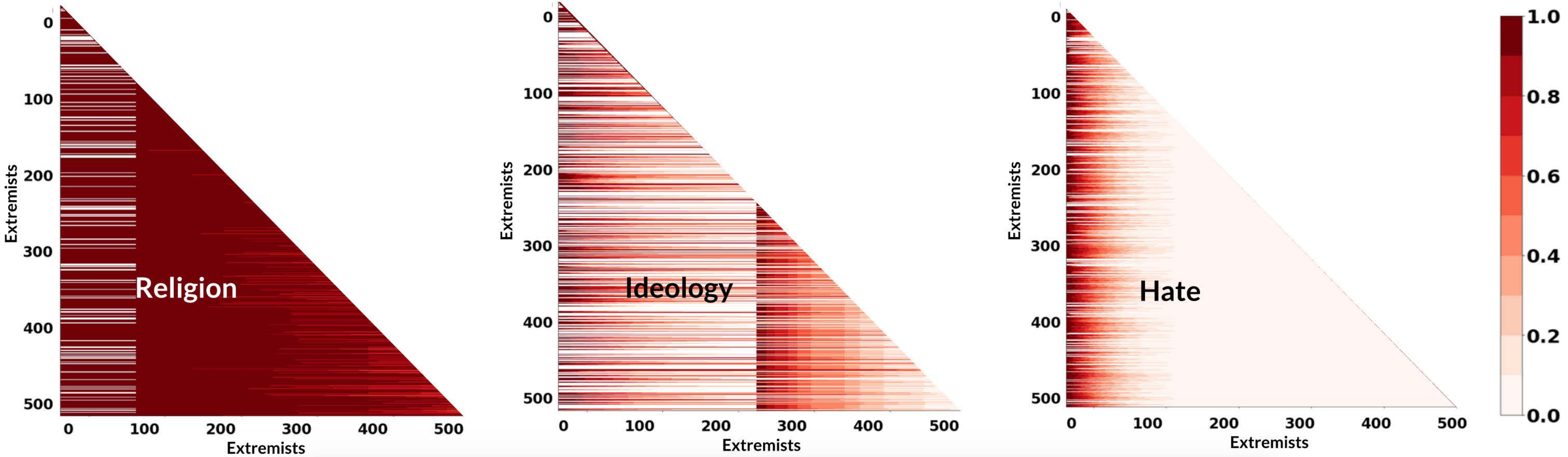}
  \end{center}
  \caption{\footnotesize Similarity between extremist  users based on a single dimension (religion, ideology and hate). User id appears on the x and y axes for extremist users. Extremist users display strong similarity among themselves based on religion, while a small set of extremist users on the x-axis do now show this similarity with other extremist users on the y-axis. A cluster of users between 250 and 350 on the x-axis for ideology in (middle) and users between 0 and 50 for hate in (right)  show strong similarity with other users. On the other hand, a set of extremist users on the y-axis of (left) triangle-figure and x-axis of (right) triangle figure do not show any similarity (white cells) with the majority of other extremist users.}
  \label{fig:hmap-EE}
\end{figure}

\begin{figure}[!htbp]
  \begin{center}
    \includegraphics[width=0.80\textwidth, 
    trim=5.0cm 1.5cm 3.5cm 0.5cm]{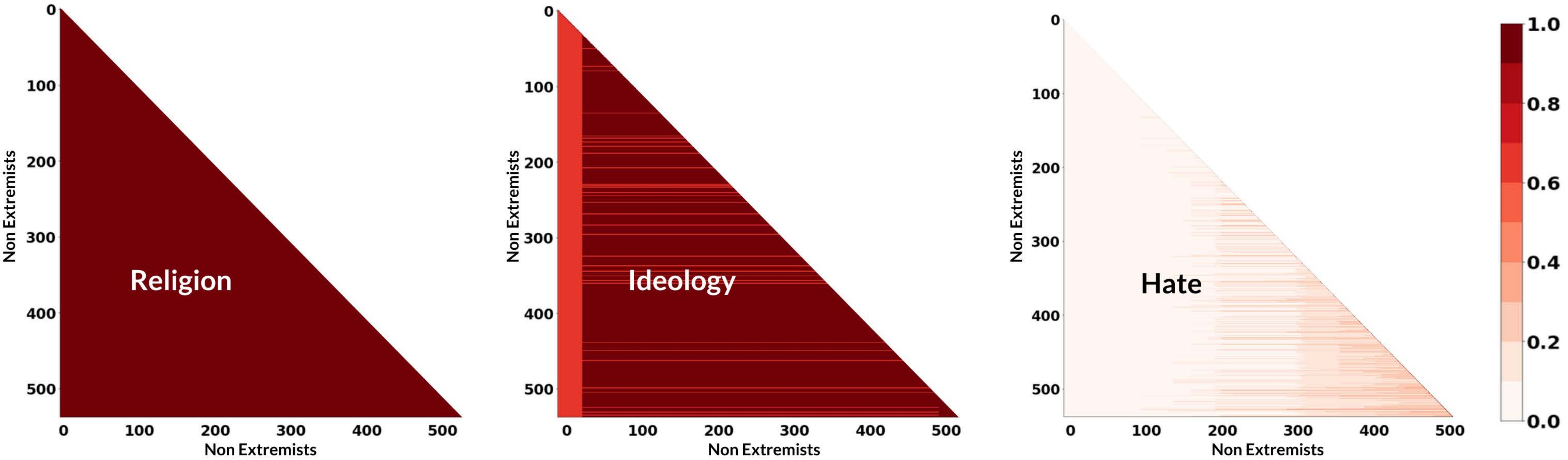}
  \end{center}
  \caption{\footnotesize Similarity between non-extremist users only based on religion(left), ideology(middle) and hate(right) dimensions. User id appears on the x and y axes for non-extremist users. Non-extremist users are strongly similar to each other based on religion as well as ideology, while they do not display similarity based on hate.}
  \label{fig:hmap-NENE}
\end{figure}

As observed in Sections \ref{sec:NG} and \ref{sec:TA}, the content of extremist and non-extremist users show strong dissimilarities in the use of language and the topics of conversations based on extremist Islamist ideology and hate, whereas they are relatively similar based on religion. Assessment of similarity between extremist and non-extremist users  reveals the contrast between these users with respect to the three contextual dimensions. Further, assessing similarity between users in each group (extremist/extremist, non-extremist/non-extremist), shows the coherence of the extremist and non-extremist datasets, and further allows us to observe potential anomalies if any exist. Hence, we provide a similarity analysis of users through comparison between  pairs of users. We utilize the embedding representations of such users created through three contextual dimension models (R, I, H) (see Section \ref{sec:CDM})   and measure the distance between them using cosine similarity. In Figures \ref{fig:hmap-NE}, \ref{fig:hmap-EE}, and \ref{fig:hmap-NENE}, the heat maps depict user similarity, where similarity values range from 0.0 to 1.0, represented using shades of red, with white being 0 and dark red being 1; therefore, the darker areas correspond to more similar users. 

Figure \ref{fig:hmap-NE} shows the similarity between extremist and non-extremist users. The pairs of extremist/non-extremist users (E-N) show strong similarity for religion in Figure \ref{fig:hmap-NE} (left), while the similarity is relatively weak for hate in Figure \ref{fig:hmap-NE} (right). Based on ideology, these pairs of users display significant similarity in Figure \ref{fig:hmap-NE} (middle) compared to hate. On the other hand, it is noteworthy that a small set of pairs for both hate and ideology have stronger similarity compared to other users.

Figure \ref{fig:hmap-EE} depicts similarity among extremist users for religion, ideology and hate contextual dimensions. In Figure \ref{fig:hmap-EE} (left), extremist users generally show strong similarity based on religion. On ideology, while a significant number of extremist users are not similar with each other, a collection of extremist users between 250 and 350 on the x-axis shows stronger similarity with other extremist users between 240 and 538 on the y-axis in Figure \ref{fig:hmap-EE} (middle). In Figure \ref{fig:hmap-EE} (right), only a small collection of extremist users between 0 and 50 on the x-axis shows strong similarity with the majority of other extremist users on the y-axis based on hate. We believe that this is due to extremist users employing different hate tactics for different targets. When we consider tweets from users that are “distant” for the hate dimension, their content looks different. For example, considering the example tweets from Table \ref{tab:examples}, one extremist user’s tweets might be emboldening hatred against "apostates" (i.e., Muslims in other countries who, in the eyes of Islamist extremist groups, have deserted Muslim ideals, ideology and religion), while another might be about hatred against “the West”. The context in each of these conversations would be different because of their target, despite the hatred being incited. It is noteworthy that a spectrum of extremist users between 0 and 100 on the x-axis representing 19\% of these extremist users of Figure \ref{fig:hmap-EE} (left) for religion, and a collection of disparate users on the y-axis of Figure \ref{fig:hmap-EE} (right) for hate, do not show any similarity (white cells) with a significant number of other extremist users on the y-axis. \textit{This implies that the extremist user dataset might contain outliers or mislabeled users as extremist.}

Figure \ref{fig:hmap-NENE} depicts similarity among non-extremist users for religion and ideology dimensions. In Figure \ref{fig:hmap-NENE} (left) and (middle), non-extremist users show strong similarity based on religion and ideology respectively. Note that the darker shade of red does represent the similarity of two users, but not relatedness of their content to ideology. Hence, strongly similar non-extremist users based on ideology might still have low relatedness to specific ideological content. In Figure \ref{fig:hmap-NENE} (right), non-extremist users do not show similarity with each other based on hate. 
\paragraph{\textbf{Observations:}} Through this exploratory analysis, we make the following observations that guide our modeling approach: (i) The content of extremist users heavily contains language, topical and contextual features from Islamist extremist ideology, religion of Islam, and hate speech. (ii) While extremist and non-extremist users are similar in their appeal to religion, they differ in their appeal to Islamist extremist ideology and hate. This might be because they employ different ideological and hate tactics for their targets. (iii) A small subset of extremist users are dissimilar to other extremist users for religion (see Figure \ref{fig:hmap-EE}), implying that the extremist user dataset may contain \emph{likely outlier users}. Hence, we pursue a set of experiments to examine the presence of likely outliers and their identification in the extremist dataset, as described in Section \ref{sec:IO}

\section{Method for Modeling Extremist and Non-Extremist Users}
\label{sec:method}
In this section, we explain our modeling approach informed by our observations in Section \ref{sec:eda}. We first examine the existence of likely outliers using a chain of techniques. After we identify and remove likely outlier users in the extremist user dataset, we perform imputation to deal with sparse representations of users for the three contextual dimensions. Finally, we create and evaluate our models employing various combinations of contextual dimensions.

\subsection{Identification of Outliers}
\label{sec:IO}
In Section \ref{sec:US}, we have seen that the extremist dataset might contain anomalous users (potentially non-extremist), whom we called \emph{likely outliers}. As described in Section \ref{sec:D}, sole expertise in the Arabic language or the problem of online abusive behavior would not suffice for reliable labeling process, as such a complex problem requires deep domain knowledge and expertise. Therefore, we suspect that the likely outliers that we have observed in the extremist dataset result from the lack of knowledge and expertise in the problem of Islamist extremism. Especially, given the immense sensitivity and related security implications of the problem, we undertake a further in-depth analysis that includes hierarchical clustering and statistical analysis, followed by validation by our co-author domain expert in the field of religious extremism. 
 
\begin{figure}[t]
  \centering
    \includegraphics[width=0.85\textwidth, 
    trim=5.0cm 1.5cm 4.0cm 1.5cm]{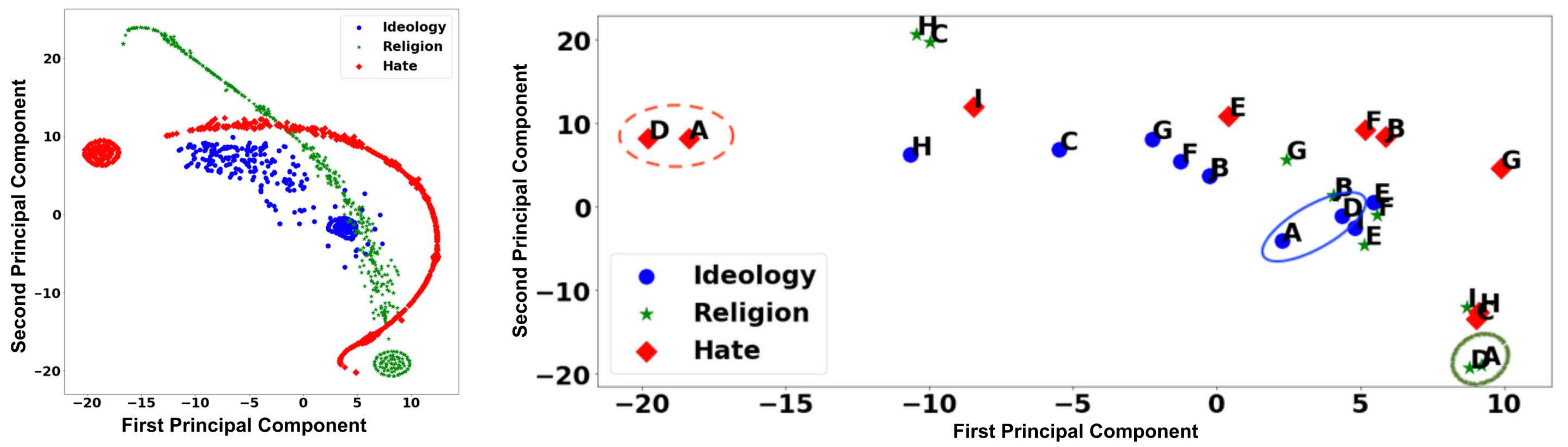}
  \caption{\footnotesize Placement of representations of users in contextual dimensions of religion, ideology and hate in 2-D space using t-SNE (Best seen in color). In (left), which overlays three representations in different coordinates, users show similarity on their spread over the space based on hate and religion, while a small cluster of users fall far from others. (right) provides a closer look over random samples of 20 users on the 2-d space. Users A and D (circled) are close to each other for all contextual dimensions, while they potentially form an outlier cluster of users for hate and religion.}
  \label{fig:US-2D}
  \vspace{-1em}
\end{figure}

To visualize potential dissimilarity and the presence of likely outliers among the extremist users with respect to contextual dimensions, we plot representations of extremist users in a two dimensional space in Figure \ref{fig:US-2D} (left), using T-distributed Stochastic Neighbor Embedding (t-SNE) \cite{maaten2008visualizing}. It demonstrates such dissimilarity, the spread of users over the space, and existence of likely outlier users for the three contextual dimensions. When we stretch the patterns of the spread of users for hate and religion, it provides strong similarity, while a circling set of users for both contextual dimensions forming the cluster of likely outliers. To ensure that the users in these small clusters are same users for both contextual dimensions, we have picked 10 random users from the extremist user dataset and placed their representations in each contextual dimension on the 2-D space. As shown in Figure \ref{fig:US-2D} (right), users \textbf{A and D} fall far from other users forming an outlier cluster for religion and hate contextual dimensions. We have repeated this procedure with different sets of random 10 users multiple times, and found that the users in these small clusters are same, which confirms our observations from Figure \ref{fig:US-2D} (left). Therefore, we must identify these likely outliers before creating our models. 

\paragraph{\textbf{Hierarchical Clustering:}} To identify the likely outliers  in the extremist user dataset, we perform an unsupervised hierarchical density based clustering (HDBC) \cite{campello2013density} over the 538 users in the extremist dataset for each contextual dimension. HDBC forms clusters based on the euclidean distance between the users with respect to their representations for each of the contextual dimensions of religion, ideology and hate. HDBC clustering reveals two main clusters, one of which forms the majority of users, the \emph{Likely Extremist} users, while the small cluster of users is the \emph{Likely Outliers}. 
Figure \ref{fig:hdbc} shows the distribution of users over the two clusters, where the y axis represents the percentage of users in each cluster for the dimensions of religion, ideology and hate.

\paragraph{\textbf{Statistical Analysis:}} From the HDBC clustering, we identified 99 (18\%), 48 (9\%) and 141 (26\%) users in the extremist dataset, clustered as \emph{likely outliers} for \emph{religion, ideology and hate contextual dimensions,} respectively. To confirm a clear separation between these two clusters along with its statistical significance, we performed a non-parametric Mann-Whitney U-test for each contextual dimension. 
Table \ref{tab:TPE} shows the U-statistics (U-stats) and their p-values along with the effect size. While all analyses reveal a significant difference, all effects are significant with moderate effect sizes.



The effect size for ideology is slightly higher than that for the hate and the religion, implying that ideology is more effective in clustering. This outcome suggests that the variance between the content of the two clusters of users based on each dimension (especially ideology) is high; hence, the users in these two clusters are significantly different from each other. 

\vspace{1em}
\begin{minipage}{\textwidth}
\begin{minipage}[b]{0.48\textwidth}
 \centering
    \includegraphics[width=0.65\textwidth, 
    trim=5.0cm 2.5cm 3.5cm 0.5cm]{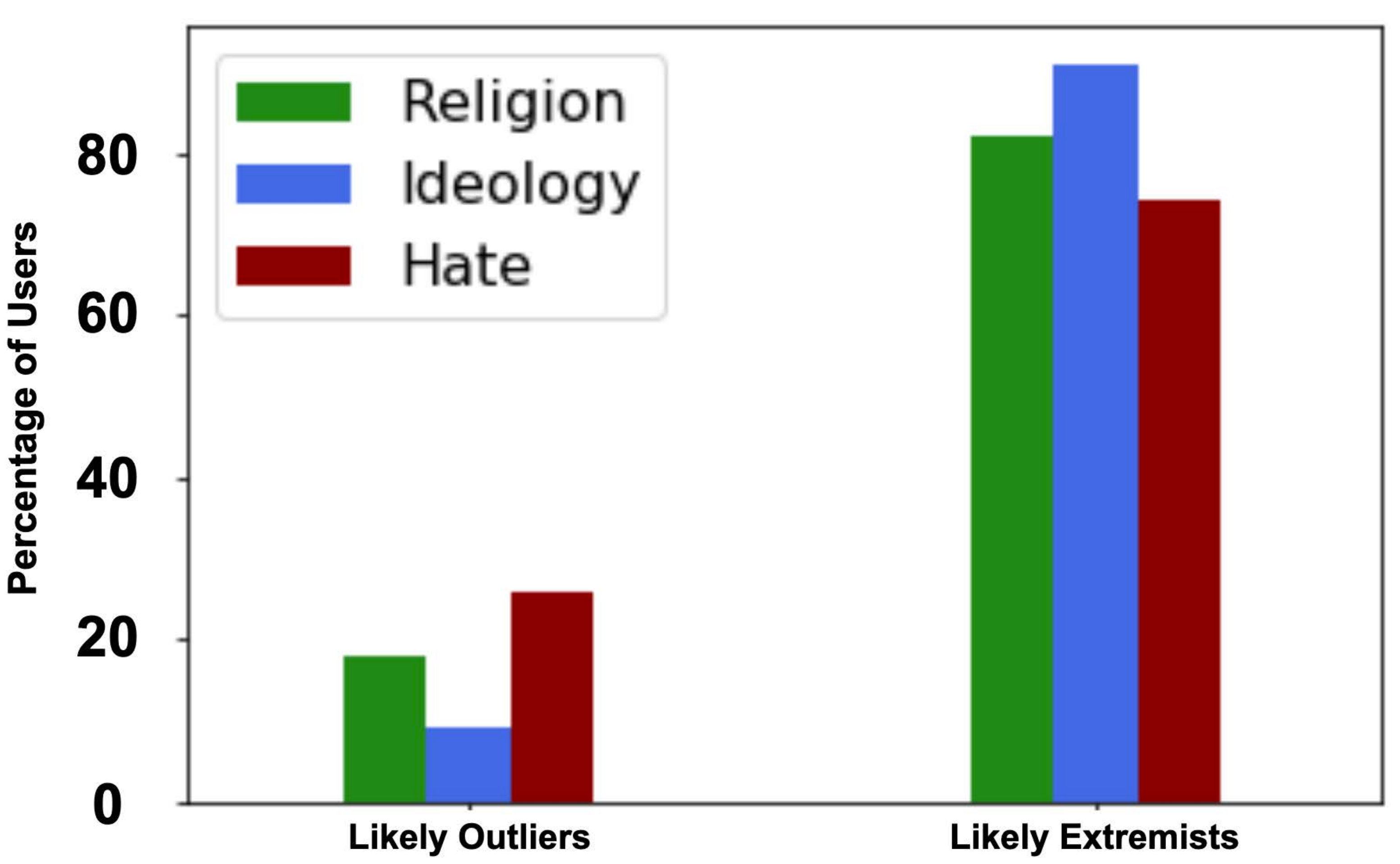}
  \captionof{figure}{\footnotesize Based on the HDBC clustering algorithm, we obtain two main clusters, and we call the majority cluster as “Likely Extremist” and the small cluster as “Likely Outliers”. The y-axis represents the percentage of users for each cluster. 141 hate, 99 religion, 48 ideology.}
  \label{fig:hdbc}
\end{minipage}
\hfill
\begin{minipage}[b]{0.50\textwidth}
\begin{tabular}{p{1.2cm}p{1.0cm}p{0.9cm}p{1cm}p{0.9cm}}
    \toprule[2pt]
     \textbf{\footnotesize{Dimension}} & \textbf{\footnotesize{U -stats}} & \textbf{\footnotesize{z-score}} & \textbf{\footnotesize{p-value}} & \textbf{\footnotesize{Effect Size}} \\ \midrule[2pt]
     Religion & 5049 & 12.08 & 0.0027 & 0.53 \\
     Ideology & 9566 & 13.95 & 0.001 & 0.61 \\
     Hate & 8178 & 12.4 & 0.0016 & 0.54 \\
      \bottomrule[2pt]
\end{tabular}
\captionof{table}{\footnotesize Non-parametric Mann-Whitney U-test between extremist users using their content representations from different dimensions. While outliers and likely extremist users differ on all dimensions, the did not differ quite as much on religion.} 
\label{tab:TPE}
\end{minipage}
\end{minipage}

\paragraph{\textbf{Validation:}} We created a random sample of 76 users comprising 15\% of the extremist dataset, to validate the two clusters for identified likely outliers and likely extremists. Our co-author domain expert annotated these users as \emph{likely outliers, and likely extremist}. We obtained a kappa score \cite{mchugh2012interrater} of 0.82, (69 correct and 7 incorrect matches). 

Lastly, upon the validation of outliers by our co-author domain expert, we obtained the set of 49 outlier users in the extremist dataset. Content of the outlier users contains the following frequent terms: \emph{marriage, Allah, bonded, silence, Islam leaders, Berjaya hilarious, cake, miss mit, kemaren, Quran, Khuda, prophet, Muhammad, Ahmad}. We found that these outlier users different from other extremist users, and they are most likely non-extremist users. Keeping these outlier users in the extremist user dataset will cause the model to \emph{unfairly} classify non-extremist users as extremist, which will create serious implications in a real world scenario. Hence, we remove this set of outlier users from the dataset to be used in our modeling phase, yielding an extremist dataset with 489 users. 

\vspace{-0.5em}
\subsection{Imputation for Sparse Representations}
\label{sec:ISP}
Users on social media often use slang terms and informal language rather than the archaic language used in religious and ideological resources. Moreover, some users use hateful language mixed with religious terms, concepts, and topics in their content, while they do not share information related to Islamist extremist ideology. On the other hand, some users mostly share ideological content while they neither share religious content nor use hate speech. This situation creates sparse content of users in any single dimension, translating to sparse embedding vectors (in some cases zero vectors). We have identified 148 users who had relatively sparse contextual content for at least one of the three dimensions. 

Missing values in a dataset is a common problem, and statistical approaches have been developed to approximate such missing components for numerical data. Similarly, natural language processing addresses this problem through, such as out of vocabulary (OOV) \cite{blunsom2016proceedings} words found in domain-specific applications \cite{sarma2018domain}. 

\begin{wrapfigure}{R}{0.5\textwidth}
\vspace{-2.5em}
\begin{minipage}{0.5\textwidth}
  \begin{algorithm}[H] 
    \caption{Imputation for Extremist Users (E)}
    \label{alg:loop}
    \begin{algorithmic}[1]
    \Require{$\Psi$, $\mathbf{\hat{U}_E}$, $\mathbf{U_E}$, $\mathbf{\Gamma_E}$} \Comment{Set of dimensions, Users with sparse vectors, All users, Topic model, respectively.}
    \Statex
    \Function{imputation}{$\Psi$, $\mathbf{\hat{U}_E}$, $\mathbf{U_E}$, $\mathbf{\Gamma_E}$} 
    \For{$d$ in $\Psi$}    
        \For{$\hat{u}^d$ in $\mathbf{\hat{U}_E}$}
            \State $\widetilde{u}^d$ $\gets$ $\max_{u^d \in \mathbf{U^d_E}}$ $\Big\{\frac{\Gamma_E(\hat{u}^d) \cap \Gamma_E(u^d)  }{\Gamma_E(\hat{u}^d) \cup \Gamma_E(u^d)}\Big\}$
            \State $\vec{\mathcal{V}}(\hat{u}^d)$ $\gets$ $\vec{\mathcal{V}}({\widetilde{u}^d})$ \Comment{Equation \ref{eq:1}}
        \EndFor
    \EndFor
    \EndFunction
    \end{algorithmic}
    \label{alg:1}
    \end{algorithm}
\end{minipage}
\vspace{-1.2em}
\end{wrapfigure}

We utilize LDA to address the sparse representation problem by using “imputation” based on the topical similarity of of user content \cite{zhang2014exploit}. We train two LDA models (see Section \ref{sec:TA}), one for the extremist dataset ($\mathbf{\Gamma_E}$) and another for non-extremist dataset ($\mathbf{\Gamma_N}$). We generate topics from the content of each user. We compute topical similarity between users with sparse representations $\mathbf{\hat{U}_E}$ and users with dense representations ($\mathbf{U^R_E}$, $\mathbf{U^I_E}$, $\mathbf{U^H_E}$). After we assess topical similarity, we create topics that will represent $\mathbf{\hat{U}_E}$, taking the ratio of intersection topics between $\mathbf{\hat{U}_E}$ and $\mathbf{U^R_E}$ or $\mathbf{U^I_E}$ or $\mathbf{U^H_E}$ over the union of the topics between between $\mathbf{\hat{U}_E}$ and $\mathbf{U^R_E}$ or $\mathbf{U^I_E}$ or $\mathbf{U^H_E}$, and a set of topics that represents a user with sparse representation is denoted as $\widetilde{u}^d$ \cite{aletras2014measuring}. Then, we generate the embeddings of $\widetilde{u}^d$, resulting the imputed user representations, in line 5 of Algorithm \ref{alg:1}. To assess the impact of this approach on performance of our modeling approach, we create our models with and without imputed data, and report performance improvements in Section \ref{sec:results}.

\vspace{-0.8em}
\subsection{Modeling}
We develop models employing different combinations (uni-bi-tri-dimensional) of the three contextual  dimensions (religion, ideology, hate) to identify the best possible representation of users for classification \cite{kursuncu2018s}, and determine the effectiveness and contribution of each dimension. We generate vector representations of users (489 extremist users after removal of outliers and 538 non-extremists users) for each dimension, and concatenate them to create uni-dimensional, bi-dimensional and tri-dimensional models. Uni-dimensional models include only the representation for one contextual dimension, bi-dimensional models include two dimensions and tri-dimensional model include all three dimensions. Then, for the bi-dimensional and tri-dimensional models, we perform SVD to reduce the dimensionality of the vector after concatenation, down to 300. Further, we perform our experiments developing models with and without the imputed representations to assess the effectiveness of imputation for sparse representation of users (see Section \ref{sec:ISP}). For models without imputation, we eliminate users with sparse representation, which correspond to removing 148 users from our dataset. 

Our hypothesis is that a model with three dimensions will create more coherent representations of users leading to improvements in the performance of classification. To test our hypothesis, we create and compare models that include uni-dimensional (R, I, H), bi-dimensional (IH, RI, RH) and tri-dimensional (RIH) models with and without imputation, apart from the baseline model (see below).  

Since the existing related work \cite{ferrara2016predicting, fernandez2018understanding} has utilized Random Forest (RF) and Naive Bayes (NB) algorithms, we employ the same algorithms for a fair comparison. As we identified and removed 49 outliers from the extremist dataset (see Section \ref{sec:IO}), we start with 1027 users with imputation and 879 users without imputation, and then create a hold-out dataset of 300 users. We perform training using stratified 6-fold cross-validation. In Section \ref{sec:results}, we report the results for our modeling approach, discuss possible implications, and provide comparison with our baseline. We have chosen the state-of-the-art baseline model defined in \cite{fernandez2018understanding} which is grounded in social science models of radicalization. Note that this is a modeling comparison using our dataset, described in Section \ref{sec:D}. They use a frequency-based weighting scheme with a NB model for dichotomous classification over two levels (micro and meso). As we were unable to secure  their proprietary resources (i.e., lexicon) used by \cite{fernandez2018understanding}, we made our best effort at replicating their approach on our dataset, for a fair comparison.

\section{Results}
\label{sec:results}
\begin{figure}[!htbp]
  \begin{center}
    \includegraphics[width=0.25\textwidth,
    trim=7.0cm 5.5cm 7.5cm 4.5cm, angle=-90]{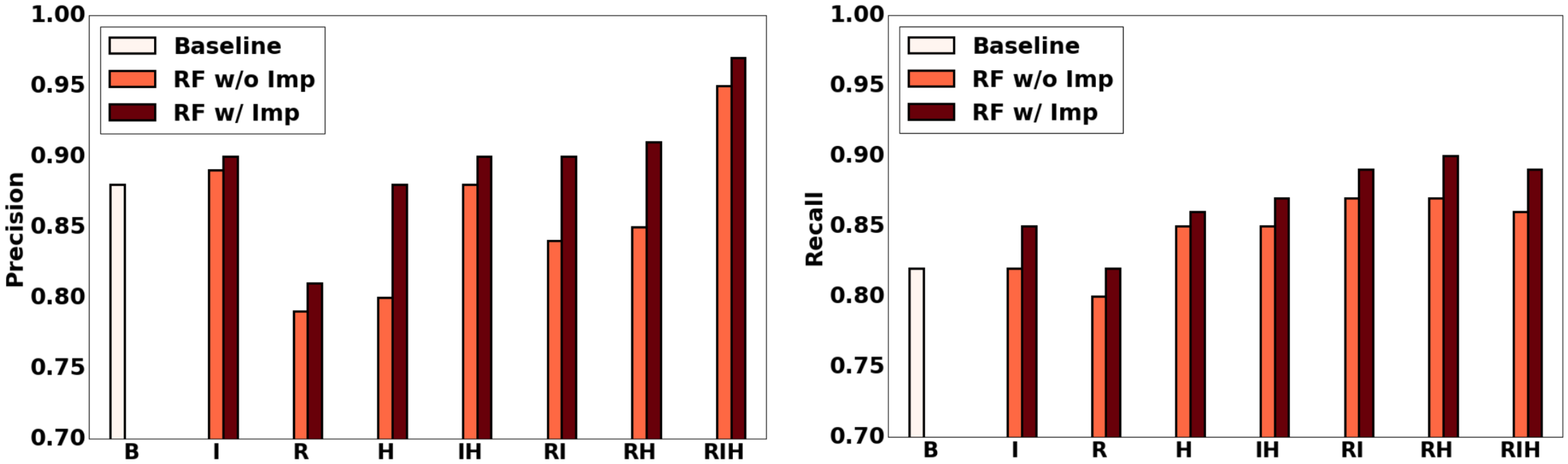}
  \end{center}
  \caption{\footnotesize Precision and Recall of the models using Random Forest (RF) with and without imputation, based on different combinations of contextual dimensions of religion (R), ideology (I) and hate (H). In general, the models with imputation outperform other models without imputation. In (a), the tri-dimensional (RIH) model provides the best performance with 0.97 precision, and in (b) the bi-dimensional (RH) model provides best recall of 0.9, followed by the RIH model with 0.89.}
  \label{fig:PRplot}
 \vspace{-0.5em}
\end{figure}

In this section, we report the results on performance of the models we created using precision, recall, F1-score and AUC metrics. From Table \ref{tab:results} and Figure \ref{fig:PRplot}, the baseline model has a precision of 0.88, recall of 0.82 and F1-score of 0.84 using a feature size of 23K based on the frequency of unigrams \cite{fernandez2018understanding}.

As shown in Table \ref{tab:results}, \emph{RF models with imputation} outperform others in precision, recall and F1-score. Uni-dimensional models with imputation achieve a precision of 0.90 using only ideology, recall of 0.86 using only hate and F1-score of 0.87 with ideology and hate each. We observe that the bi-dimensional model with the combination of religion and hate provides better precision, recall and F1-score of 0.91, 0.90 and 0.91, respectively. The combination of ideology dimension individually with the other two dimensions achieves a modest improvement (2.2\%) in precision, and a greater improvement (8.5\%) in recall. The inclusion of ideology improves recall reducing false negatives, and improving the identification of extremist users. 

When we combine the three contextual dimensions, our tri-dimensional model achieves best performance with a precision of 0.97, a recall of 0.89 and an F1-score of 0.93, with imputation. Our tri-dimension model improves the precision, recall and F1-score over the baseline model by 9.3\%, 7.9\%, 10.7\%, respectively. On the other hand, the tri-dimensional model (RIH) improves precision and F1-score over the bi-dimensional model (RH) by 6.6\% and 2.2\% respectively, at the expense of a decrease of 1.1\% in recall. Reducing misclassification of non-extremist users creates a minor decrease in recall, which can be considered a trade-off for a significant improvement in precision, with respect to a large set of non-extremist users. This trade-off between the tri-dimensional (RIH) and bi-dimensional (RH) models in a real world application, translates as follows: while more ($\sim$1\%) extremist users  are mislabeled as non-extremist (false negatives), less ($\sim$6\%) non-extremist users are mislabeled as extremist (false positives). \emph{As misclassification of non-extremist users can have significant implications in a large-scale application where non-extremists vastly outnumber extremists, the higher precision reduces potential social discrimination.}

\begin{table}[t]
\begin{center}
\begin{tabular}[t]{p{2.2cm}ccccccc}
    \toprule[3pt]
    \textbf{Dimension} & \textbf{Algorithm} & \multicolumn{2}{c}{\textbf{Precision}} & \multicolumn{2}{c}{\textbf{Recall}} & \multicolumn{2}{c}{\textbf{F1-Score}} \\
    \cmidrule{3-8}
    & & \textbf{w/o Imp} & \textbf{w/ Imp} & \textbf{w/o Imp} & \textbf{w/ Imp} & \textbf{w/o Imp} & \textbf{w/ Imp} \\ \midrule[2pt]
    \textbf{Baseline} & NB & 0.88 & -- & 0.82 & -- & 0.84 & -- \\ \midrule[2pt]
    \rowcolor[gray]{.8}
    Ideology (I) & RF & 0.89 & 0.90 & 0.82 & 0.85 & 0.85 & 0.87 \\
    \rowcolor[gray]{.8}
    Religion (R) & RF & 0.79 & 0.81 & 0.80 & 0.82 & 0.80 & 0.81 \\
    \rowcolor[gray]{.8}
    Hate (H) & RF & 0.84 & 0.88 & 0.85 & 0.86 & 0.85 & 0.87 \\ \midrule
    I & NB & 0.80 & 0.88 & 0.71 & 0.75 & 0.75 & 0.81 \\ 
    R & NB & 0.70 & 0.79 & 0.71 & 0.74 & 0.75 & 0.76 \\
    H & NB & 0.79 & 0.80 & 0.81 & 0.85 & 0.80 & 0.83 \\ \midrule[2pt]
    \rowcolor[gray]{.8}
    I+H & RF & 0.88 & 0.90 & 0.85 & 0.87 & 0.86 & 0.89 \\
    \rowcolor[gray]{.8}
    I+R & RF & 0.84 & 0.90 & \textbf{0.87} & 0.89 & 0.86 & 0.89 \\
    \rowcolor[gray]{.8}
    R+H & RF & 0.85 & 0.91 & \textbf{0.87} & \textbf{0.90} & 0.86 & 0.91 \\ \midrule
    I+H & NB & 0.88 & 0.88 & 0.83 & 0.85 & 0.85 & 0.86 \\
    I+R & NB & 0.86 & 0.89 & 0.81 & 0.87 & 0.83 & 0.88 \\ 
    R+H & NB & 0.81 & 0.92 & 0.80 & 0.84 & 0.80 & 0.88 \\ \midrule[2pt]
    \rowcolor[gray]{.8}
    \textbf{R+I+H }& \textbf{RF} & \textbf{0.95} & \textbf{0.97} & 0.86 & 0.89 & \textbf{0.91} & \textbf{0.93} \\ \midrule
    R+I+H & NB & 0.90 & 0.91 & 0.82 & 0.82 & 0.86 & 0.87 \\
    \bottomrule[3pt]
\end{tabular}
\end{center}
\caption{\footnotesize Results of the uni-bi-tri-dimensional models with and without imputation (Imp). The models without imputation were created based on 879 users after the removal of 49 identified outlier users and 148 users with sparse representations as described in section \ref{sec:IO}. The models with imputation were created based on the 1027 users after the removal of the 49 identified outlier users.}
\label{tab:results}
\vspace{-2.5em}
\end{table}

To better illustrate the diagnostic ability of these models, we plot ROC curves and compute AUC scores where we can examine the performance of the models at different thresholds with respect to true positive rate (TPR) and false positive rate (FPR). Higher TPR with lower FPR indicates better performance. When a model approaches 1.0 of TPR, the corresponding lower FPR at this point compared to other models signifies better performance. 

\begin{figure}[t]
  \begin{center}
    \includegraphics[width=0.8\textwidth, 
    trim=5.0cm 1.5cm 1.5cm 0.5cm]{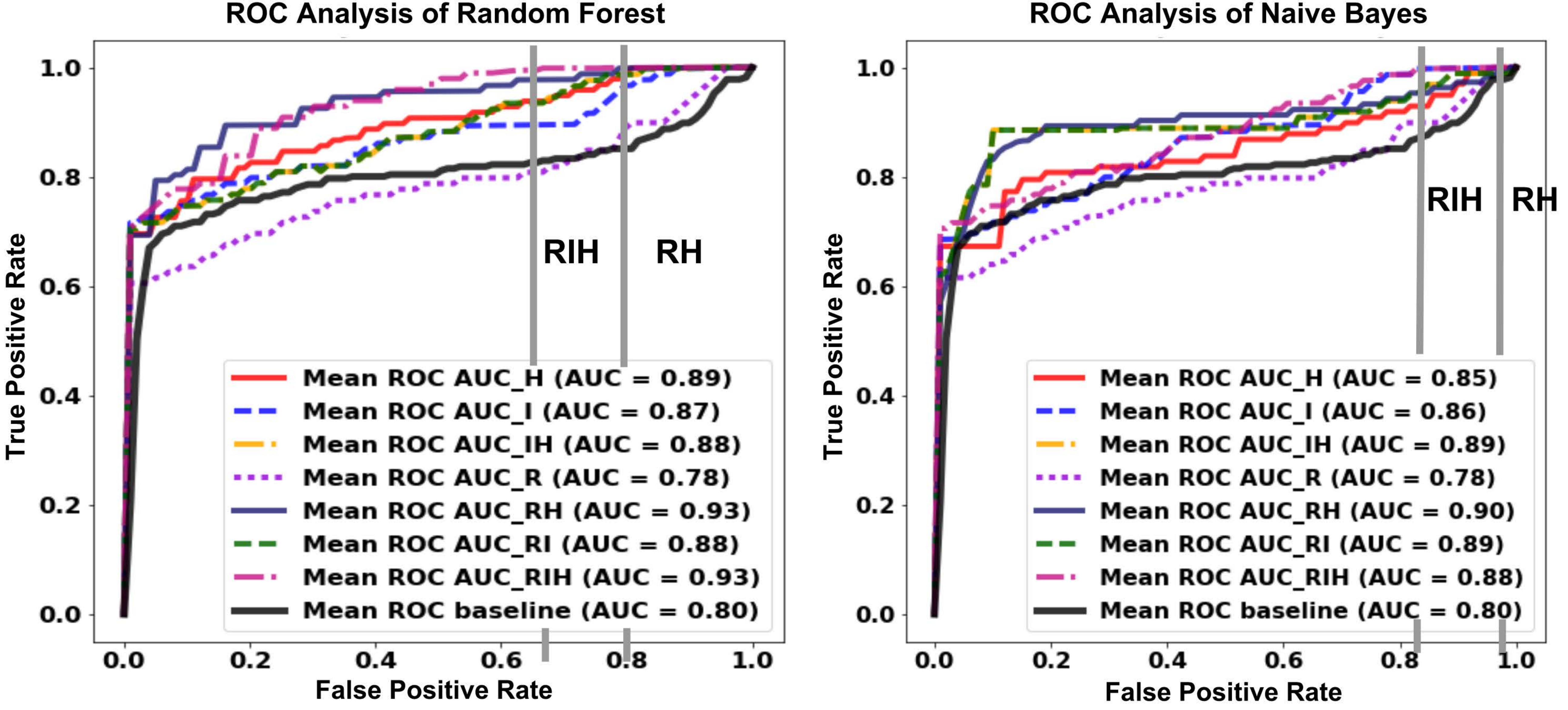}
  \end{center}
  \caption{\footnotesize ROC curves and AUC scores of uni-bi-tri-dimensional models with RF and NB (Best seen in color). The RIH tri-dimensional and RH bi-dimensional RF models outperform other models with an AUC score of 0.93. The ROC curve for RIH converges to 1.0 at TPR earlier than the RH model, providing better precision. Note: The baseline approach involves only NB and was not tested  with RF.}
  \label{fig:roc_curves}
\vspace{-2em}
\end{figure}

Figure \ref{fig:roc_curves} shows ROC curves for RF and NB models with different dimensions. Using RF, our tri-dimensional model along with the bi-dimensional model RH, achieves the best performance with AUC of 0.93, improving upon the baseline by 16.3\%. Further, the ROC curve of the tri-dimensional RF model converges to 1.0 of true positive rate (TPR) at the false positive rate (FPR) of 0.65, whereas the tri-dimensional NB model reaches 1.0 of TPR at FPR of 0.82, implying a 10.9\% gain in precision. 


Moreover, using NB, our models with multiple contextual dimensions, except the uni-dimensional model with religion, are outperforming the baseline in AUC (80\%) by up to 12.5\%. This shows the effectiveness of our approach with contextual dimensions of Religion, Ideology and Hate, using the same classifier, namely Naive Bayes. However, the bi-dimensional model RH provides the best performance, with AUC of 0.90, followed by the two bi-dimensional models with RI and IH. While the bi-dimensional model outperforms the tri-dimensional model, the ROC curve of the tri-dimensional NB model converges to 1.0 of TPR at FPR of 0.83, whereas the bi-dimensional model reaches the closest point to 1.0 of TPR at FPR of 0.97, implying a 9.8\% gain in precision. 

\paragraph{\textbf{Key Insights:}} (i) Ideology and hate dimensions are often coupled with religious concepts in the content of extremist users. The inclusion of all three contextual dimensions displays the best performance compared to other models in terms of precision. Specifically, this improvement is important because it significantly reduces potential security implications of a possible deployment of this model; thus, reducing the likelihood of an unfair mistreatment towards non-extremist individuals, in a real world application. (ii) Considering that all three dimensions performing well, different extremist users employ diverse strategies to effectively cover broader set of followers at different levels of radicalization. Specifically, given that each of the three contextual dimensions plays different roles in different levels of radicalization, tri-dimensional model captures nuances as well as linguistic and semantic cues better with respect to different density of these contexts throughout the radicalization process. (iii) The contextual dimensions of religion and hate had more power in classification, suggesting that the extremist users were often using religious content along with hate language. This maybe because they use religious concepts, events, places and historic figures, to justify their hatred towards their targets, such as "apostates" or "the West", as in the examples 2 and 3 in Table \ref{tab:examples}; they could encourage their followers to commit acts of \emph{violence}.



\section{Conclusion}
\label{sec:conclusion}
Using a principled, multi-dimensional approach to the analysis of Islamist extremist content as defined in social science, we excluded likely outlier non-extremist users to develop a robust classifier with improved precision. The success of our method highlights the limitations of more superficial, manual approaches to the identification of extremist users. 
Furthermore, it suggests a radicalization process over time through a careful contextual metering approach involving religion, Islamist extremist ideology and hate conversations, with fine granularity.

We improved upon the state-of-the-art in automated classification using the three contextual dimensions of Islamist extremism on social media and learned three domain-specific embedding models for interpreting content shared by the users. Overall, our comprehensive approach achieved 10.2\%, 8.5\% and  10.7\% improvement in precision, recall and F1-score, respectively, over a competing baseline. We make the dataset and domain specific corpora for the three dimensions available upon request for research and reproducibility purposes.

\paragraph{\textbf{Limitations and Future work:}} The limited number of labeled instances available for training may fail to track the changing nature of concepts and relationships. We plan to address the dynamic nature of this problem in our future work. Specifically, as the past data may not be representative of the future dynamic changes, we will explore the use of domain-specific knowledge of the radicalization process and progression to make our analysis less fragile.

%
\section*{Acknowledgement}
We acknowledge partial support from the National Science Foundation (NSF) award CNS-1513721: “Context-Aware Harassment Detection on Social Media". C. Castillo was partially supported by La Caixa (LCF) project (LCF/PR/PR16/11110009). D. Achilov was supported in part by the University of Notre Dame (UND) Global Religion Research Initiative (GRRI) grant through Templeton Religion Trust (Grant ID: TRT0118). Any opinions, conclusions or recommendations expressed in this material are those of the authors and do not necessarily reflect the views of the NSF, LCF, or UND.

We also thank reviewers of the CSCW 2019 for their constructive feedback that greatly improved the presentation of this work.


%
\bibliographystyle{ACM-Reference-Format}
\bibliography{Reference}


\begin{thebibliography}{61}


\ifx \showCODEN    \undefined \def \showCODEN     #1{\unskip}     \fi
\ifx \showDOI      \undefined \def \showDOI       #1{#1}\fi
\ifx \showISBNx    \undefined \def \showISBNx     #1{\unskip}     \fi
\ifx \showISBNxiii \undefined \def \showISBNxiii  #1{\unskip}     \fi
\ifx \showISSN     \undefined \def \showISSN      #1{\unskip}     \fi
\ifx \showLCCN     \undefined \def \showLCCN      #1{\unskip}     \fi
\ifx \shownote     \undefined \def \shownote      #1{#1}          \fi
\ifx \showarticletitle \undefined \def \showarticletitle #1{#1}   \fi
\ifx \showURL      \undefined \def \showURL       {\relax}        \fi
\providecommand\bibfield[2]{#2}
\providecommand\bibinfo[2]{#2}
\providecommand\natexlab[1]{#1}
\providecommand\showeprint[2][]{arXiv:#2}

\bibitem[\protect\citeauthoryear{Achilov and Sen}{Achilov and Sen}{2017}]%
        {achilov2017got}
\bibfield{author}{\bibinfo{person}{Dilshod Achilov} {and}
  \bibinfo{person}{Sedat Sen}.} \bibinfo{year}{2017}\natexlab{}.
\newblock \showarticletitle{Got political Islam? Are politically moderate
  Muslims really different from radicals?}
\newblock \bibinfo{journal}{\emph{International Political Science Review}}
  (\bibinfo{year}{2017}).
\newblock


\bibitem[\protect\citeauthoryear{Agarwal and Sureka}{Agarwal and
  Sureka}{2014}]%
        {agarwal2014focused}
\bibfield{author}{\bibinfo{person}{Swati Agarwal} {and} \bibinfo{person}{Ashish
  Sureka}.} \bibinfo{year}{2014}\natexlab{}.
\newblock \showarticletitle{A focused crawler for mining hate and extremism
  promoting videos on YouTube.}. In \bibinfo{booktitle}{\emph{Proceedings of
  the 25th ACM conference on Hypertext and social media}}.
\newblock


\bibitem[\protect\citeauthoryear{Agarwal and Sureka}{Agarwal and
  Sureka}{2015}]%
        {agarwal2015using}
\bibfield{author}{\bibinfo{person}{Swati Agarwal} {and} \bibinfo{person}{Ashish
  Sureka}.} \bibinfo{year}{2015}\natexlab{}.
\newblock \showarticletitle{Using knn and svm based one-class classifier for
  detecting online radicalization on twitter}. In
  \bibinfo{booktitle}{\emph{International Conference on Distributed Computing
  and Internet Technology}}.
\newblock


\bibitem[\protect\citeauthoryear{Agarwal and Sureka}{Agarwal and
  Sureka}{2016}]%
        {agarwal2016spider}
\bibfield{author}{\bibinfo{person}{Swati Agarwal} {and} \bibinfo{person}{Ashish
  Sureka}.} \bibinfo{year}{2016}\natexlab{}.
\newblock \showarticletitle{Spider and the flies: Focused crawling on tumblr to
  detect hate promoting communities}.
\newblock \bibinfo{journal}{\emph{arXiv preprint arXiv:1603.09164}}
  (\bibinfo{year}{2016}).
\newblock


\bibitem[\protect\citeauthoryear{Agarwal, Sureka, and Goyal}{Agarwal
  et~al\mbox{.}}{2015}]%
        {agarwal2015open}
\bibfield{author}{\bibinfo{person}{Swati Agarwal}, \bibinfo{person}{Ashish
  Sureka}, {and} \bibinfo{person}{Vikram Goyal}.}
  \bibinfo{year}{2015}\natexlab{}.
\newblock \showarticletitle{Open source social media analytics for intelligence
  and security informatics applications}. In
  \bibinfo{booktitle}{\emph{International Conference on Big Data Analytics}}.
\newblock


\bibitem[\protect\citeauthoryear{Alava, Frau-Meigs, and Hassan}{Alava
  et~al\mbox{.}}{2017}]%
        {alava2017youth}
\bibfield{author}{\bibinfo{person}{S{\'e}raphin Alava}, \bibinfo{person}{Divina
  Frau-Meigs}, {and} \bibinfo{person}{Ghayda Hassan}.}
  \bibinfo{year}{2017}\natexlab{}.
\newblock \bibinfo{booktitle}{\emph{Youth and violent extremism on social
  media: mapping the research}}.
\newblock


\bibitem[\protect\citeauthoryear{Aletras and Stevenson}{Aletras and
  Stevenson}{2014}]%
        {aletras2014measuring}
\bibfield{author}{\bibinfo{person}{Nikolaos Aletras} {and}
  \bibinfo{person}{Mark Stevenson}.} \bibinfo{year}{2014}\natexlab{}.
\newblock \showarticletitle{Measuring the similarity between automatically
  generated topics}. In \bibinfo{booktitle}{\emph{Proceedings of the 14th
  Conference of the European Chapter of the Association for Computational
  Linguistics, volume 2: Short Papers}}. \bibinfo{pages}{22--27}.
\newblock


\bibitem[\protect\citeauthoryear{Anwar and Abulaish}{Anwar and
  Abulaish}{2015}]%
        {anwar2015ranking}
\bibfield{author}{\bibinfo{person}{Tarique Anwar} {and}
  \bibinfo{person}{Muhammad Abulaish}.} \bibinfo{year}{2015}\natexlab{}.
\newblock \showarticletitle{Ranking radically influential web forum users}.
\newblock \bibinfo{journal}{\emph{IEEE Transactions on Information Forensics
  and Security}} (\bibinfo{year}{2015}).
\newblock


\bibitem[\protect\citeauthoryear{Arpinar, Kursuncu, and Achilov}{Arpinar
  et~al\mbox{.}}{2016}]%
        {arpinar2016social}
\bibfield{author}{\bibinfo{person}{I~Budak Arpinar}, \bibinfo{person}{Ugur
  Kursuncu}, {and} \bibinfo{person}{Dilshod Achilov}.}
  \bibinfo{year}{2016}\natexlab{}.
\newblock \showarticletitle{Social media analytics to identify and counter
  islamist extremism: Systematic detection, evaluation, and challenging of
  extremist narratives online}. In \bibinfo{booktitle}{\emph{2016 International
  Conference on Collaboration Technologies and Systems (CTS)}}. IEEE,
  \bibinfo{pages}{611--612}.
\newblock


\bibitem[\protect\citeauthoryear{Ashcroft, Fisher, Kaati, Omer, and
  Prucha}{Ashcroft et~al\mbox{.}}{2015}]%
        {ashcroft2015detecting}
\bibfield{author}{\bibinfo{person}{Michael Ashcroft}, \bibinfo{person}{Ali
  Fisher}, \bibinfo{person}{Lisa Kaati}, \bibinfo{person}{Enghin Omer}, {and}
  \bibinfo{person}{Nico Prucha}.} \bibinfo{year}{2015}\natexlab{}.
\newblock \showarticletitle{Detecting jihadist messages on twitter}. In
  \bibinfo{booktitle}{\emph{2015 European Intelligence and Security Informatics
  Conference}}.
\newblock


\bibitem[\protect\citeauthoryear{Athiwaratkun, Wilson, and
  Anandkumar}{Athiwaratkun et~al\mbox{.}}{2018}]%
        {athiwaratkun2018probabilistic}
\bibfield{author}{\bibinfo{person}{Ben Athiwaratkun},
  \bibinfo{person}{Andrew~Gordon Wilson}, {and} \bibinfo{person}{Anima
  Anandkumar}.} \bibinfo{year}{2018}\natexlab{}.
\newblock \showarticletitle{Probabilistic fasttext for multi-sense word
  embeddings}.
\newblock \bibinfo{journal}{\emph{arXiv preprint arXiv:1806.02901}}
  (\bibinfo{year}{2018}).
\newblock


\bibitem[\protect\citeauthoryear{Awan}{Awan}{2017}]%
        {awan2017cyber}
\bibfield{author}{\bibinfo{person}{Imran Awan}.}
  \bibinfo{year}{2017}\natexlab{}.
\newblock \showarticletitle{Cyber-extremism: Isis and the power of social
  media}.
\newblock \bibinfo{journal}{\emph{Society}} \bibinfo{volume}{54},
  \bibinfo{number}{2} (\bibinfo{year}{2017}), \bibinfo{pages}{138--149}.
\newblock


\bibitem[\protect\citeauthoryear{Badawy and Ferrara}{Badawy and
  Ferrara}{2018}]%
        {badawy2018rise}
\bibfield{author}{\bibinfo{person}{Adam Badawy} {and} \bibinfo{person}{Emilio
  Ferrara}.} \bibinfo{year}{2018}\natexlab{}.
\newblock \showarticletitle{The rise of jihadist propaganda on social
  networks}.
\newblock \bibinfo{journal}{\emph{Journal of Computational Social Science}}
  (\bibinfo{year}{2018}).
\newblock


\bibitem[\protect\citeauthoryear{Berger and Morgan}{Berger and Morgan}{2015}]%
        {berger2015isis}
\bibfield{author}{\bibinfo{person}{Jonathon~M Berger} {and}
  \bibinfo{person}{Jonathon Morgan}.} \bibinfo{year}{2015}\natexlab{}.
\newblock \showarticletitle{The ISIS Twitter Census: Defining and describing
  the population of ISIS supporters on Twitter}.
\newblock \bibinfo{journal}{\emph{The Brookings Project on US Relations with
  the Islamic World}} (\bibinfo{year}{2015}).
\newblock


\bibitem[\protect\citeauthoryear{Blunsom, Cho, Cohen, Grefenstette, Hermann,
  Rimell, Weston, and Yih}{Blunsom et~al\mbox{.}}{2016}]%
        {blunsom2016proceedings}
\bibfield{author}{\bibinfo{person}{Phil Blunsom}, \bibinfo{person}{Kyunghyun
  Cho}, \bibinfo{person}{Shay Cohen}, \bibinfo{person}{Edward Grefenstette},
  \bibinfo{person}{Karl~Moritz Hermann}, \bibinfo{person}{Laura Rimell},
  \bibinfo{person}{Jason Weston}, {and} \bibinfo{person}{Scott Wen-tau Yih}.}
  \bibinfo{year}{2016}\natexlab{}.
\newblock \showarticletitle{Proceedings of the 1st Workshop on Representation
  Learning for NLP}. In \bibinfo{booktitle}{\emph{Proceedings of the 1st
  Workshop on Representation Learning for NLP}}.
\newblock


\bibitem[\protect\citeauthoryear{Bojanowski, Grave, Joulin, and
  Mikolov}{Bojanowski et~al\mbox{.}}{2017}]%
        {bojanowski2017enriching}
\bibfield{author}{\bibinfo{person}{Piotr Bojanowski}, \bibinfo{person}{Edouard
  Grave}, \bibinfo{person}{Armand Joulin}, {and} \bibinfo{person}{Tomas
  Mikolov}.} \bibinfo{year}{2017}\natexlab{}.
\newblock \showarticletitle{Enriching word vectors with subword information}.
\newblock \bibinfo{journal}{\emph{Transactions of the Association for
  Computational Linguistics}}  \bibinfo{volume}{5} (\bibinfo{year}{2017}),
  \bibinfo{pages}{135--146}.
\newblock


\bibitem[\protect\citeauthoryear{Bouma}{Bouma}{2009}]%
        {bouma2009normalized}
\bibfield{author}{\bibinfo{person}{Gerlof Bouma}.}
  \bibinfo{year}{2009}\natexlab{}.
\newblock \showarticletitle{Normalized (pointwise) mutual information in
  collocation extraction}.
\newblock \bibinfo{journal}{\emph{Proceedings of GSCL}} (\bibinfo{year}{2009}),
  \bibinfo{pages}{31--40}.
\newblock


\bibitem[\protect\citeauthoryear{Boutin, Chauzal, Dorsey, Jegerings, Paulussen,
  Pohl, Reed, and Zavagli}{Boutin et~al\mbox{.}}{2016}]%
        {boutin2016foreign}
\bibfield{author}{\bibinfo{person}{B{\'e}r{\'e}nice Boutin},
  \bibinfo{person}{Gr{\'e}gory Chauzal}, \bibinfo{person}{Jessica Dorsey},
  \bibinfo{person}{Marjolein Jegerings}, \bibinfo{person}{Christophe
  Paulussen}, \bibinfo{person}{Johanna Pohl}, \bibinfo{person}{Alastair Reed},
  {and} \bibinfo{person}{Sofia Zavagli}.} \bibinfo{year}{2016}\natexlab{}.
\newblock \showarticletitle{The foreign fighters phenomenon in the European
  Union}.
\newblock \bibinfo{journal}{\emph{Profiles, threats \& policies. The
  International Centre for Counter-Terrorism--The Hague. doi}}
  (\bibinfo{year}{2016}).
\newblock


\bibitem[\protect\citeauthoryear{Bowman-Grieve and Conway}{Bowman-Grieve and
  Conway}{2012}]%
        {bowman2012exploring}
\bibfield{author}{\bibinfo{person}{Lorraine Bowman-Grieve} {and}
  \bibinfo{person}{Maura Conway}.} \bibinfo{year}{2012}\natexlab{}.
\newblock \showarticletitle{Exploring the form and function of dissident Irish
  Republican online discourses}.
\newblock \bibinfo{journal}{\emph{Media, War \& Conflict}}
  (\bibinfo{year}{2012}).
\newblock


\bibitem[\protect\citeauthoryear{Bunt}{Bunt}{2003}]%
        {bunt2003islam}
\bibfield{author}{\bibinfo{person}{Gary~R Bunt}.}
  \bibinfo{year}{2003}\natexlab{}.
\newblock \bibinfo{booktitle}{\emph{Islam in the digital age: E-jihad, online
  fatwas and cyber Islamic environments}}.
\newblock \bibinfo{publisher}{Pluto Press}.
\newblock


\bibitem[\protect\citeauthoryear{Campello, Moulavi, and Sander}{Campello
  et~al\mbox{.}}{2013}]%
        {campello2013density}
\bibfield{author}{\bibinfo{person}{Ricardo~JGB Campello},
  \bibinfo{person}{Davoud Moulavi}, {and} \bibinfo{person}{J{\"o}rg Sander}.}
  \bibinfo{year}{2013}\natexlab{}.
\newblock \showarticletitle{Density-based clustering based on hierarchical
  density estimates}. In \bibinfo{booktitle}{\emph{Pacific-Asia conference on
  knowledge discovery and data mining}}.
\newblock


\bibitem[\protect\citeauthoryear{Cano~Basave, He, Liu, and Zhao}{Cano~Basave
  et~al\mbox{.}}{2013}]%
        {cano2013weakly}
\bibfield{author}{\bibinfo{person}{Amparo~Elizabeth Cano~Basave},
  \bibinfo{person}{Yulan He}, \bibinfo{person}{Kang Liu}, {and}
  \bibinfo{person}{Jun Zhao}.} \bibinfo{year}{2013}\natexlab{}.
\newblock \showarticletitle{A weakly supervised bayesian model for violence
  detection in social media}.
\newblock  (\bibinfo{year}{2013}).
\newblock


\bibitem[\protect\citeauthoryear{Chen, Weber, and Okulicz-Kozaryn}{Chen
  et~al\mbox{.}}{2014}]%
        {chen2014us}
\bibfield{author}{\bibinfo{person}{Lu Chen}, \bibinfo{person}{Ingmar Weber},
  {and} \bibinfo{person}{Adam Okulicz-Kozaryn}.}
  \bibinfo{year}{2014}\natexlab{}.
\newblock \showarticletitle{US religious landscape on Twitter}. In
  \bibinfo{booktitle}{\emph{International Conference on Social Informatics}}.
\newblock


\bibitem[\protect\citeauthoryear{Cook}{Cook}{2015}]%
        {cook2015understanding}
\bibfield{author}{\bibinfo{person}{David Cook}.}
  \bibinfo{year}{2015}\natexlab{}.
\newblock \bibinfo{booktitle}{\emph{Understanding jihad}}.
\newblock \bibinfo{publisher}{Univ of California Press}.
\newblock


\bibitem[\protect\citeauthoryear{Davidson, Warmsley, Macy, and Weber}{Davidson
  et~al\mbox{.}}{2017}]%
        {hateoffensive}
\bibfield{author}{\bibinfo{person}{Thomas Davidson}, \bibinfo{person}{Dana
  Warmsley}, \bibinfo{person}{Michael Macy}, {and} \bibinfo{person}{Ingmar
  Weber}.} \bibinfo{year}{2017}\natexlab{}.
\newblock \showarticletitle{Automated Hate Speech Detection and the Problem of
  Offensive Language}. In \bibinfo{booktitle}{\emph{Proceedings of the 11th
  International AAAI Conference on Web and Social Media}}
  \emph{(\bibinfo{series}{ICWSM '17})}. \bibinfo{pages}{512--515}.
\newblock


\bibitem[\protect\citeauthoryear{De~Leede, Haupfleisch, Korolkova, and
  Natter}{De~Leede et~al\mbox{.}}{2017}]%
        {de2017radicalisation}
\bibfield{author}{\bibinfo{person}{Seran De~Leede}, \bibinfo{person}{Renate
  Haupfleisch}, \bibinfo{person}{Katja Korolkova}, {and}
  \bibinfo{person}{Monika Natter}.} \bibinfo{year}{2017}\natexlab{}.
\newblock \bibinfo{booktitle}{\emph{Radicalisation and Violent Extremism-Focus
  on Women: How Women Become Radicalised, and how to Empower Them to Prevent
  Radicalisation}}.
\newblock


\bibitem[\protect\citeauthoryear{Fernandez and Alani}{Fernandez and
  Alani}{2018}]%
        {fernandez2018contextual}
\bibfield{author}{\bibinfo{person}{Miriam Fernandez} {and}
  \bibinfo{person}{Harith Alani}.} \bibinfo{year}{2018}\natexlab{}.
\newblock \showarticletitle{Contextual Semantics for Radicalisation Detection
  on Twitter}.
\newblock  (\bibinfo{year}{2018}).
\newblock


\bibitem[\protect\citeauthoryear{Fernandez, Asif, and Alani}{Fernandez
  et~al\mbox{.}}{2018}]%
        {fernandez2018understanding}
\bibfield{author}{\bibinfo{person}{Miriam Fernandez}, \bibinfo{person}{Moizzah
  Asif}, {and} \bibinfo{person}{Harith Alani}.}
  \bibinfo{year}{2018}\natexlab{}.
\newblock \showarticletitle{Understanding the roots of radicalisation on
  twitter}. In \bibinfo{booktitle}{\emph{Proceedings of the 10th ACM Conference
  on Web Science}}.
\newblock


\bibitem[\protect\citeauthoryear{Ferrara, Wang, Varol, Flammini, and
  Galstyan}{Ferrara et~al\mbox{.}}{2016}]%
        {ferrara2016predicting}
\bibfield{author}{\bibinfo{person}{Emilio Ferrara}, \bibinfo{person}{Wen-Qiang
  Wang}, \bibinfo{person}{Onur Varol}, \bibinfo{person}{Alessandro Flammini},
  {and} \bibinfo{person}{Aram Galstyan}.} \bibinfo{year}{2016}\natexlab{}.
\newblock \showarticletitle{Predicting online extremism, content adopters, and
  interaction reciprocity}. In \bibinfo{booktitle}{\emph{International
  conference on social informatics}}.
\newblock


\bibitem[\protect\citeauthoryear{Frampton, Fisher, Prucha, and
  Petraeus}{Frampton et~al\mbox{.}}{2017}]%
        {frampton2017new}
\bibfield{author}{\bibinfo{person}{Martyn Frampton}, \bibinfo{person}{Ali
  Fisher}, \bibinfo{person}{Nico Prucha}, {and} \bibinfo{person}{David~H
  Petraeus}.} \bibinfo{year}{2017}\natexlab{}.
\newblock \bibinfo{booktitle}{\emph{The new Netwar: Countering extremism
  online}}.
\newblock \bibinfo{publisher}{Policy Exchange}.
\newblock


\bibitem[\protect\citeauthoryear{Gaur, Alambo, Sain, Kursuncu, Thirunarayan,
  Kavuluru, Sheth, Welton, and Pathak}{Gaur et~al\mbox{.}}{2019}]%
        {gaur2019knowledge}
\bibfield{author}{\bibinfo{person}{Manas Gaur}, \bibinfo{person}{Amanuel
  Alambo}, \bibinfo{person}{Joy~Prakash Sain}, \bibinfo{person}{Ugur Kursuncu},
  \bibinfo{person}{Krishnaprasad Thirunarayan}, \bibinfo{person}{Ramakanth
  Kavuluru}, \bibinfo{person}{Amit Sheth}, \bibinfo{person}{Randy Welton},
  {and} \bibinfo{person}{Jyotishman Pathak}.} \bibinfo{year}{2019}\natexlab{}.
\newblock \showarticletitle{Knowledge-aware assessment of severity of suicide
  risk for early intervention}. In \bibinfo{booktitle}{\emph{The World Wide Web
  Conference}}. ACM, \bibinfo{pages}{514--525}.
\newblock


\bibitem[\protect\citeauthoryear{Gaur, Kursuncu, Alambo, Sheth, Daniulaityte,
  Thirunarayan, and Pathak}{Gaur et~al\mbox{.}}{2018}]%
        {gaur2018let}
\bibfield{author}{\bibinfo{person}{Manas Gaur}, \bibinfo{person}{Ugur
  Kursuncu}, \bibinfo{person}{Amanuel Alambo}, \bibinfo{person}{Amit Sheth},
  \bibinfo{person}{Raminta Daniulaityte}, \bibinfo{person}{Krishnaprasad
  Thirunarayan}, {and} \bibinfo{person}{Jyotishman Pathak}.}
  \bibinfo{year}{2018}\natexlab{}.
\newblock \showarticletitle{Let Me Tell You About Your Mental Health!:
  Contextualized Classification of Reddit Posts to DSM-5 for Web-based
  Intervention}. In \bibinfo{booktitle}{\emph{Proceedings of the 27th ACM
  International Conference on Information and Knowledge Management}}. ACM,
  \bibinfo{pages}{753--762}.
\newblock


\bibitem[\protect\citeauthoryear{Hafez and Mullins}{Hafez and Mullins}{2015}]%
        {hafez2015radicalization}
\bibfield{author}{\bibinfo{person}{Mohammed Hafez} {and}
  \bibinfo{person}{Creighton Mullins}.} \bibinfo{year}{2015}\natexlab{}.
\newblock \showarticletitle{The radicalization puzzle: A theoretical synthesis
  of empirical approaches to homegrown extremism}.
\newblock \bibinfo{journal}{\emph{Studies in Conflict \& Terrorism}}
  (\bibinfo{year}{2015}).
\newblock


\bibitem[\protect\citeauthoryear{Helfstein}{Helfstein}{2012}]%
        {helfstein2012edges}
\bibfield{author}{\bibinfo{person}{Scott Helfstein}.}
  \bibinfo{year}{2012}\natexlab{}.
\newblock \bibinfo{booktitle}{\emph{Edges of radicalization: Ideas, individuals
  and networks in violent extremism}}.
\newblock \bibinfo{type}{{T}echnical {R}eport}. \bibinfo{institution}{MILITARY
  ACADEMY WEST POINT NY COMBATING TERRORISM CENTER}.
\newblock


\bibitem[\protect\citeauthoryear{Hussain and Saltman}{Hussain and
  Saltman}{2014}]%
        {hussain2014jihad}
\bibfield{author}{\bibinfo{person}{Ghaffar Hussain} {and}
  \bibinfo{person}{Erin~Marie Saltman}.} \bibinfo{year}{2014}\natexlab{}.
\newblock \bibinfo{booktitle}{\emph{Jihad trending: A comprehensive analysis of
  online extremism and how to counter it}}.
\newblock


\bibitem[\protect\citeauthoryear{Kaati, Omer, Prucha, and Shrestha}{Kaati
  et~al\mbox{.}}{2015}]%
        {kaati2015detecting}
\bibfield{author}{\bibinfo{person}{Lisa Kaati}, \bibinfo{person}{Enghin Omer},
  \bibinfo{person}{Nico Prucha}, {and} \bibinfo{person}{Amendra Shrestha}.}
  \bibinfo{year}{2015}\natexlab{}.
\newblock \showarticletitle{Detecting multipliers of jihadism on twitter}. In
  \bibinfo{booktitle}{\emph{2015 IEEE International Conference on Data Mining
  Workshop (ICDMW)}}.
\newblock


\bibitem[\protect\citeauthoryear{Kursuncu, Gaur, Lokala, Illendula,
  Thirunarayan, Daniulaityte, Sheth, and Arpinar}{Kursuncu
  et~al\mbox{.}}{2018}]%
        {kursuncu2018s}
\bibfield{author}{\bibinfo{person}{Ugur Kursuncu}, \bibinfo{person}{Manas
  Gaur}, \bibinfo{person}{Usha Lokala}, \bibinfo{person}{Anurag Illendula},
  \bibinfo{person}{Krishnaprasad Thirunarayan}, \bibinfo{person}{Raminta
  Daniulaityte}, \bibinfo{person}{Amit Sheth}, {and} \bibinfo{person}{I~Budak
  Arpinar}.} \bibinfo{year}{2018}\natexlab{}.
\newblock \showarticletitle{What's ur Type? Contextualized Classification of
  User Types in Marijuana-Related Communications Using Compositional Multiview
  Embedding}. In \bibinfo{booktitle}{\emph{2018 IEEE/WIC/ACM International
  Conference on Web Intelligence (WI)}}. IEEE, \bibinfo{pages}{474--479}.
\newblock


\bibitem[\protect\citeauthoryear{Kursuncu, Gaur, Lokala, Thirunarayan, Sheth,
  and Arpinar}{Kursuncu et~al\mbox{.}}{2019}]%
        {kursuncu2019predictive}
\bibfield{author}{\bibinfo{person}{Ugur Kursuncu}, \bibinfo{person}{Manas
  Gaur}, \bibinfo{person}{Usha Lokala}, \bibinfo{person}{Krishnaprasad
  Thirunarayan}, \bibinfo{person}{Amit Sheth}, {and} \bibinfo{person}{I~Budak
  Arpinar}.} \bibinfo{year}{2019}\natexlab{}.
\newblock \showarticletitle{Predictive Analysis on Twitter: Techniques and
  Applications}.
\newblock In \bibinfo{booktitle}{\emph{Emerging Research Challenges and
  Opportunities in Computational Social Network Analysis and Mining}}.
  \bibinfo{publisher}{Springer}, \bibinfo{pages}{67--104}.
\newblock


\bibitem[\protect\citeauthoryear{Loza}{Loza}{2007}]%
        {loza2007psychology}
\bibfield{author}{\bibinfo{person}{Wagdy Loza}.}
  \bibinfo{year}{2007}\natexlab{}.
\newblock \showarticletitle{The psychology of extremism and terrorism: A
  Middle-Eastern perspective}.
\newblock \bibinfo{journal}{\emph{Aggression and Violent Behavior}}
  (\bibinfo{year}{2007}).
\newblock


\bibitem[\protect\citeauthoryear{Maaten and Hinton}{Maaten and Hinton}{2008}]%
        {maaten2008visualizing}
\bibfield{author}{\bibinfo{person}{Laurens van~der Maaten} {and}
  \bibinfo{person}{Geoffrey Hinton}.} \bibinfo{year}{2008}\natexlab{}.
\newblock \showarticletitle{Visualizing data using t-SNE}.
\newblock \bibinfo{journal}{\emph{Journal of machine learning research}}
  (\bibinfo{year}{2008}).
\newblock


\bibitem[\protect\citeauthoryear{McHugh}{McHugh}{2012}]%
        {mchugh2012interrater}
\bibfield{author}{\bibinfo{person}{Mary~L McHugh}.}
  \bibinfo{year}{2012}\natexlab{}.
\newblock \showarticletitle{Interrater reliability: the kappa statistic}.
\newblock \bibinfo{journal}{\emph{Biochemia medica: Biochemia medica}}
  (\bibinfo{year}{2012}).
\newblock


\bibitem[\protect\citeauthoryear{Meleagrou-Hitchens, Hughes, and
  Clifford}{Meleagrou-Hitchens et~al\mbox{.}}{2018}]%
        {meleagrou2018travelers}
\bibfield{author}{\bibinfo{person}{Alexander Meleagrou-Hitchens},
  \bibinfo{person}{Seamus Hughes}, {and} \bibinfo{person}{Bennett Clifford}.}
  \bibinfo{year}{2018}\natexlab{}.
\newblock \bibinfo{booktitle}{\emph{The travelers: American jihadists in Syria
  and Iraq}}.
\newblock


\bibitem[\protect\citeauthoryear{Mikolov, Chen, Corrado, and Dean}{Mikolov
  et~al\mbox{.}}{2013a}]%
        {mikolov2013efficient}
\bibfield{author}{\bibinfo{person}{Tomas Mikolov}, \bibinfo{person}{Kai Chen},
  \bibinfo{person}{Greg Corrado}, {and} \bibinfo{person}{Jeffrey Dean}.}
  \bibinfo{year}{2013}\natexlab{a}.
\newblock \showarticletitle{Efficient estimation of word representations in
  vector space}.
\newblock \bibinfo{journal}{\emph{arXiv preprint arXiv:1301.3781}}
  (\bibinfo{year}{2013}).
\newblock


\bibitem[\protect\citeauthoryear{Mikolov, Sutskever, Chen, Corrado, and
  Dean}{Mikolov et~al\mbox{.}}{2013b}]%
        {mikolov2013distributed}
\bibfield{author}{\bibinfo{person}{Tomas Mikolov}, \bibinfo{person}{Ilya
  Sutskever}, \bibinfo{person}{Kai Chen}, \bibinfo{person}{Greg~S Corrado},
  {and} \bibinfo{person}{Jeff Dean}.} \bibinfo{year}{2013}\natexlab{b}.
\newblock \showarticletitle{Distributed representations of words and phrases
  and their compositionality}. In \bibinfo{booktitle}{\emph{Advances in neural
  information processing systems}}.
\newblock


\bibitem[\protect\citeauthoryear{Pennington, Socher, and Manning}{Pennington
  et~al\mbox{.}}{2014}]%
        {pennington2014glove}
\bibfield{author}{\bibinfo{person}{Jeffrey Pennington},
  \bibinfo{person}{Richard Socher}, {and} \bibinfo{person}{Christopher
  Manning}.} \bibinfo{year}{2014}\natexlab{}.
\newblock \showarticletitle{Glove: Global vectors for word representation}. In
  \bibinfo{booktitle}{\emph{Proceedings of the 2014 conference on empirical
  methods in natural language processing (EMNLP)}}.
  \bibinfo{pages}{1532--1543}.
\newblock


\bibitem[\protect\citeauthoryear{Qin, Zhou, Reid, Lai, and Chen}{Qin
  et~al\mbox{.}}{2007}]%
        {qin2007analyzing}
\bibfield{author}{\bibinfo{person}{Jialun Qin}, \bibinfo{person}{Yilu Zhou},
  \bibinfo{person}{Edna Reid}, \bibinfo{person}{Guanpi Lai}, {and}
  \bibinfo{person}{Hsinchun Chen}.} \bibinfo{year}{2007}\natexlab{}.
\newblock \showarticletitle{Analyzing terror campaigns on the internet:
  Technical sophistication, content richness, and Web interactivity}.
\newblock \bibinfo{journal}{\emph{International Journal of Human-Computer
  Studies}} \bibinfo{volume}{65}, \bibinfo{number}{1} (\bibinfo{year}{2007}),
  \bibinfo{pages}{71--84}.
\newblock


\bibitem[\protect\citeauthoryear{Rowe and Saif}{Rowe and Saif}{2016}]%
        {rowe2016mining}
\bibfield{author}{\bibinfo{person}{Matthew Rowe} {and} \bibinfo{person}{Hassan
  Saif}.} \bibinfo{year}{2016}\natexlab{}.
\newblock \showarticletitle{Mining pro-ISIS radicalisation signals from social
  media users}. In \bibinfo{booktitle}{\emph{Tenth International AAAI
  Conference on Web and Social Media}}.
\newblock


\bibitem[\protect\citeauthoryear{Saif, Dickinson, Kastler, Fernandez, and
  Alani}{Saif et~al\mbox{.}}{2017}]%
        {saif2017semantic}
\bibfield{author}{\bibinfo{person}{Hassan Saif}, \bibinfo{person}{Thomas
  Dickinson}, \bibinfo{person}{Leon Kastler}, \bibinfo{person}{Miriam
  Fernandez}, {and} \bibinfo{person}{Harith Alani}.}
  \bibinfo{year}{2017}\natexlab{}.
\newblock \showarticletitle{A semantic graph-based approach for radicalisation
  detection on social media}. In \bibinfo{booktitle}{\emph{European semantic
  web conference}}.
\newblock


\bibitem[\protect\citeauthoryear{Sarma, Liang, and Sethares}{Sarma
  et~al\mbox{.}}{2018}]%
        {sarma2018domain}
\bibfield{author}{\bibinfo{person}{Prathusha~K Sarma}, \bibinfo{person}{Yingyu
  Liang}, {and} \bibinfo{person}{William~A Sethares}.}
  \bibinfo{year}{2018}\natexlab{}.
\newblock \showarticletitle{Domain Adapted Word Embeddings for Improved
  Sentiment Classification}.
\newblock \bibinfo{journal}{\emph{arXiv preprint arXiv:1805.04576}}
  (\bibinfo{year}{2018}).
\newblock


\bibitem[\protect\citeauthoryear{Scanlon and Gerber}{Scanlon and
  Gerber}{2014}]%
        {scanlon2014automatic}
\bibfield{author}{\bibinfo{person}{Jacob~R Scanlon} {and}
  \bibinfo{person}{Matthew~S Gerber}.} \bibinfo{year}{2014}\natexlab{}.
\newblock \showarticletitle{Automatic detection of cyber-recruitment by violent
  extremists}.
\newblock \bibinfo{journal}{\emph{Security Informatics}}
  (\bibinfo{year}{2014}).
\newblock


\bibitem[\protect\citeauthoryear{Scanlon and Gerber}{Scanlon and
  Gerber}{2015}]%
        {scanlon2015forecasting}
\bibfield{author}{\bibinfo{person}{Jacob~R Scanlon} {and}
  \bibinfo{person}{Matthew~S Gerber}.} \bibinfo{year}{2015}\natexlab{}.
\newblock \showarticletitle{Forecasting violent extremist cyber recruitment}.
\newblock \bibinfo{journal}{\emph{IEEE Transactions on Information Forensics
  and Security}} (\bibinfo{year}{2015}).
\newblock


\bibitem[\protect\citeauthoryear{Schafer}{Schafer}{2002}]%
        {schafer2002spinning}
\bibfield{author}{\bibinfo{person}{Joseph~A Schafer}.}
  \bibinfo{year}{2002}\natexlab{}.
\newblock \showarticletitle{Spinning the web of hate: Web-based hate
  propagation by extremist organizations}.
\newblock \bibinfo{journal}{\emph{Journal of Criminal Justice and Popular
  Culture}} (\bibinfo{year}{2002}).
\newblock


\bibitem[\protect\citeauthoryear{Sheth, Perera, Wijeratne, and
  Thirunarayan}{Sheth et~al\mbox{.}}{2017}]%
        {sheth2017knowledge}
\bibfield{author}{\bibinfo{person}{Amit Sheth}, \bibinfo{person}{Sujan Perera},
  \bibinfo{person}{Sanjaya Wijeratne}, {and} \bibinfo{person}{Krishnaprasad
  Thirunarayan}.} \bibinfo{year}{2017}\natexlab{}.
\newblock \showarticletitle{Knowledge will propel machine understanding of
  content: extrapolating from current examples}. In
  \bibinfo{booktitle}{\emph{Proceedings of the International Conference on Web
  Intelligence}}. ACM, \bibinfo{pages}{1--9}.
\newblock


\bibitem[\protect\citeauthoryear{Shin, Madotto, and Fung}{Shin
  et~al\mbox{.}}{2018}]%
        {shin2018interpreting}
\bibfield{author}{\bibinfo{person}{Jamin Shin}, \bibinfo{person}{Andrea
  Madotto}, {and} \bibinfo{person}{Pascale Fung}.}
  \bibinfo{year}{2018}\natexlab{}.
\newblock \showarticletitle{Interpreting Word Embeddings with Eigenvector
  Analysis}.
\newblock  (\bibinfo{year}{2018}).
\newblock


\bibitem[\protect\citeauthoryear{Srijith, Hepple, Bontcheva, and
  Preotiuc-Pietro}{Srijith et~al\mbox{.}}{2017}]%
        {srijith2017sub}
\bibfield{author}{\bibinfo{person}{PK Srijith}, \bibinfo{person}{Mark Hepple},
  \bibinfo{person}{Kalina Bontcheva}, {and} \bibinfo{person}{Daniel
  Preotiuc-Pietro}.} \bibinfo{year}{2017}\natexlab{}.
\newblock \showarticletitle{Sub-story detection in Twitter with hierarchical
  Dirichlet processes}.
\newblock \bibinfo{journal}{\emph{Information Processing \& Management}}
  (\bibinfo{year}{2017}).
\newblock


\bibitem[\protect\citeauthoryear{Sureka and Agarwal}{Sureka and
  Agarwal}{2014}]%
        {sureka2014learning}
\bibfield{author}{\bibinfo{person}{Ashish Sureka} {and} \bibinfo{person}{Swati
  Agarwal}.} \bibinfo{year}{2014}\natexlab{}.
\newblock \showarticletitle{Learning to classify hate and extremism promoting
  tweets}. In \bibinfo{booktitle}{\emph{2014 IEEE Joint Intelligence and
  Security Informatics Conference}}.
\newblock


\bibitem[\protect\citeauthoryear{Teh, Jordan, Beal, and Blei}{Teh
  et~al\mbox{.}}{2005}]%
        {teh2005sharing}
\bibfield{author}{\bibinfo{person}{Yee~W Teh}, \bibinfo{person}{Michael~I
  Jordan}, \bibinfo{person}{Matthew~J Beal}, {and} \bibinfo{person}{David~M
  Blei}.} \bibinfo{year}{2005}\natexlab{}.
\newblock \showarticletitle{Sharing clusters among related groups: Hierarchical
  Dirichlet processes}. In \bibinfo{booktitle}{\emph{Advances in neural
  information processing systems}}.
\newblock


\bibitem[\protect\citeauthoryear{Van~Hiel and Mervielde}{Van~Hiel and
  Mervielde}{2003}]%
        {van2003measurement}
\bibfield{author}{\bibinfo{person}{Alain Van~Hiel} {and} \bibinfo{person}{Ivan
  Mervielde}.} \bibinfo{year}{2003}\natexlab{}.
\newblock \showarticletitle{The measurement of cognitive complexity and its
  relationship with political extremism}.
\newblock \bibinfo{journal}{\emph{Political Psychology}}
  (\bibinfo{year}{2003}).
\newblock


\bibitem[\protect\citeauthoryear{Vidino and Hughes}{Vidino and Hughes}{2015}]%
        {vidino2015isis}
\bibfield{author}{\bibinfo{person}{Lorenzo Vidino} {and}
  \bibinfo{person}{Seamus Hughes}.} \bibinfo{year}{2015}\natexlab{}.
\newblock \bibinfo{booktitle}{\emph{ISIS in America: From retweets to Raqqa}}.
\newblock


\bibitem[\protect\citeauthoryear{Wadhwa and Bhatia}{Wadhwa and Bhatia}{2013}]%
        {wadhwa2013tracking}
\bibfield{author}{\bibinfo{person}{Pooja Wadhwa} {and} \bibinfo{person}{MPS
  Bhatia}.} \bibinfo{year}{2013}\natexlab{}.
\newblock \showarticletitle{Tracking on-line radicalization using investigative
  data mining}. In \bibinfo{booktitle}{\emph{2013 National Conference on
  Communications (NCC)}}.
\newblock


\bibitem[\protect\citeauthoryear{Zhang, Li, Chen, Zhang, and Wang}{Zhang
  et~al\mbox{.}}{2014}]%
        {zhang2014exploit}
\bibfield{author}{\bibinfo{person}{Haijun Zhang}, \bibinfo{person}{Zhoujun Li},
  \bibinfo{person}{Yan Chen}, \bibinfo{person}{Xiaoming Zhang}, {and}
  \bibinfo{person}{Senzhang Wang}.} \bibinfo{year}{2014}\natexlab{}.
\newblock \showarticletitle{Exploit latent dirichlet allocation for one-class
  collaborative filtering}. In \bibinfo{booktitle}{\emph{Proceedings of the
  23rd ACM International Conference on Conference on Information and Knowledge
  Management}}.
\newblock


\end{thebibliography}


%

\end{document}